\RequirePackage{fix-cm}
\documentclass[epjc3]{svjour3}  
\smartqed  
\RequirePackage{graphicx}
\usepackage{caption}
\usepackage{subcaption}
\usepackage{xcolor}
\usepackage{comment}
\usepackage[T1]{fontenc}

\RequirePackage[colorlinks=true
,urlcolor=blue
,anchorcolor=blue
,citecolor=blue
,filecolor=blue
,linkcolor=blue
,menucolor=blue
,linktocpage=true
,pdfa=true
]{hyperref}
\usepackage{siunitx}
\usepackage{graphicx}
\usepackage{caption}
\usepackage{subcaption}
\usepackage{xcolor}
\usepackage{comment}
\DeclareUnicodeCharacter{0301}{\hspace{-1ex}\'{ }}
\usepackage{enumitem}
\usepackage{amsmath}

\setlist[itemize]{noitemsep}

\journalname{Eur. Phys. J. Plus}
\begin{document}

\title{Comparing particle multiplicity predictions: Insights from \textsc{Pythia}, \textsc{Herwig} with the LHCb use case}


\author{Saliha Bashir\thanksref{e1,addr1}
        \and
       Agnieszka Obłąkowska-Mucha\thanksref{addr1}
       \and 
       Gloria Corti\thanksref{addr2}
}

\thankstext{e1}{e-mail: bashir@agh.edu.pl}


\institute{AGH University of Krakow \\
al. Mickiewicza 30 30-059 Krakow, Poland \label{addr1}
           \and CERN, Geneva, Switzerland \label{addr2}
}

\date{Received: date / Accepted: date}

\maketitle

\begin{abstract}
Monte Carlo Event Generators are tools for simulating outcomes of high-energy collisions and particle production in High Energy Physics (HEP), such as those conducted at the Large Hadron Collider (LHC). Two of the most widely used general-purpose event generators are \textsc{Pythia} and \textsc{Herwig}, both of which play a significant role in understanding particle production, the internal structure of the proton, and the underlying physics of interactions that take place at LHC. This paper focuses on the modelling of low-energy diffractive and minimum-bias proton-proton collisions, where soft QCD effects and MPI dominate particle production, rather than hard perturbative processes. The LHCb experiment focuses on studying heavy quark hadrons, and given its specialization in precision measurements and rare decays, the choice of event generator for simulations is crucial. In this paper, we present a comparison between simulations using \textsc{Pythia} and \textsc{Herwig} with LHCb data. Our analysis demonstrates that \textsc{Pythia} provides a more consistent agreement with experimental measurements across key observables and hence remains the preferred generator for many particle physics analyses at the LHCb experiment, ensuring its continued importance in future high-energy physics research.

\end{abstract}

\maketitle
\tableofcontents

\section{Introduction}
\label{sec:intro}
High-energy hadron collisions provide a unique opportunity to study a rich spectrum of interactions governed by the strong, electromagnetic, and weak forces. Within particle physics, the collision between two particles is referred to as an "event," producing a multitude of outgoing particles that can be detected by experimental apparatus. These events adhere to conservation laws, ensuring that the total energy, momentum, and quantum numbers of the final particles are conserved. Despite these principles, the specific properties and number of particles produced in each event can vary significantly. At the Large Hadron Collider (LHC), physicists investigate high-energy proton-proton collisions, which involve a wide array of elementary interactions between partons (quarks and gluons) and result in diverse final states, enriching our understanding of fundamental forces and particle interactions.

To study and predict the complex and stochastic nature of the processes, physicists rely on simulation tools known as event generators. These algorithms replicate complex interactions that occur during particle collisions, producing simulated events. Popular Monte Carlo event generators such as \textsc{Pythia} \cite{pythia_manual,pythia_8.1}, \textsc{Herwig} \cite{Herwig_2021,herwig73releasenote}, and Sherpa \cite{sherpa,sherpa2} are continually refined to enhance their accuracy and alignment with experimental observations.

In the LHCb collaboration, proton-proton event samples are generated primarily using \textsc{Pythia} \cite{simproject}. This paper provides an overview of both \textsc{Pythia} and \textsc{Herwig}, followed by a comparative analysis of their performance in calculating particle multiplicity in minimum-bias events, which are selected with minimal trigger requirements. Special attention is given to the modelling of low-energy diffractive interactions that represent an unbiased sample of inelastic collisions, where one or both protons break apart and produce new particles. This comparison is crucial for understanding how different models simulate particle production. To validate the predictions made by \textsc{Pythia} and \textsc{Herwig}, we compare the simulated results with experimental data as obtained with the LHCb detector \cite{rivet_V0,rivet_prompt}.

\section{General Structure of Monte Carlo Event Generators}

Monte Carlo event generators provide a modular simulation of high-energy hadronic collisions by combining (i) a perturbative short-distance calculation for the hard scattering, (ii) all-order resummation of soft/collinear radiation via parton showers, and (iii) non-perturbative models for multiple parton interactions (MPI), hadronisation and beam remnants.  The standard factorised form for the inclusive cross section is:

\begin{align}
d\sigma_{h_1h_2\rightarrow cd} &= \int_0^1 dx_1 \, dx_2 \sum_{a,b} f_{a/h_1}(x_1,\mu_F^2) \, f_{b/h_2}(x_2,\mu_F^2) \nonumber \\
&\quad \times d\hat\sigma_{ab\rightarrow cd}(\mu_F^2,\mu_R^2),
\label{factorisation}
\end{align}

For two colliding hadrons $h_1$ and $h_2$, the differential cross-section for producing partonic states $c$ and $d$ via partons $a,b \in \{q, \bar q, g\}$, 
where $x_i$ are the momentum fractions carried by the incoming partons, $\mu_F$ is the \emph{factorization scale}, which specifies the separation between long-distance physics absorbed into the parton distribution functions and the short-distance dynamics described by the partonic cross section, and 
$\mu_R$ is the \emph{renormalization scale}, which sets the scale at which the strong coupling $\alpha_s$ is evaluated after renormalization. The partonic cross-section $d\hat\sigma$ is computed at fixed order in pQCD, with the PDFs determined from global fits to experimental data \cite{aom_gen}. This modular picture and practical implementation details are discussed \cite{PDG_MC,LesHouchesGuide}.

\subsection{Hard process generation}

In a proton-proton collision, only a small fraction of interactions involve a large momentum transfer $Q^2$, producing final states that can be calculated using perturbative QCD (pQCD). Most collisions result in soft hadrons or particles travelling close to the beam direction, which are not treated as hard processes. Monte Carlo event generators focus on simulating the high-$Q^2$ core interaction, which forms the starting point for the subsequent stages of the event.

In practice, the hard process in an event generator can be generated internally (e.g., from built-in leading-order or next-to-leading-order matrix elements) or imported from dedicated fixed-order matrix element generators such as MadGraph\_aMC\text{@}NLO~\cite{madgraph,madgraph2} or Powheg Box~\cite{powheg,powheg2}. To provide a consistent description across phase space, the fixed-order calculation is combined with the parton shower using matching or merging schemes (e.g., MLM \cite{mlm} or CKKW-L \cite{CKKWatNLO}), which avoid double counting and preserve the accuracy of multi-jet final states. The resulting hard-scattering configuration serves as the input for the subsequent stages: parton showers, multi-parton interactions, and hadronisation \cite{hard_processes,matching_merging}.

\subsection{Parton showers}

A more dynamic stage of event generation concerns the fate of the incoming and outgoing partons involved in the hard collision. These coloured particles—quarks and gluons—radiate additional partons through processes analogous to electromagnetic bremsstrahlung, but governed by the non-Abelian SU(3) gauge structure of QCD. Unlike photons in QED, gluons themselves carry colour charge and can therefore emit further gluons, producing a cascade of successive branchings. This leads to an extended shower of predominantly soft and collinear radiation that gradually fills phase space. The evolution is typically simulated as a step-by-step branching process, $a \rightarrow bc$, in which a parent parton produces two daughter partons according to QCD or QED splitting functions, such as $q \rightarrow qg$, $g \rightarrow gg$, or $g \rightarrow q\bar{q}$. The branching probabilities, derived from the Dokshitzer-Gribov-Lipatov-Altarelli-Parisi (DGLAP) equations, depend on variables such as the transverse momentum scale $Q^2$ and the energy-momentum sharing between daughters. Initial-state showers evolve backwards from the hard-scattering scale toward the incoming beams, while final-state showers evolve forwards from the hard-scattering products toward lower scales. The evolution terminates when the momentum scale reaches the region where perturbation theory is no longer reliable, after which non-perturbative modelling, such as hadronisation, takes over~\cite{parton_showers,ShowerReviews}.

\subsection{Multiparton Interactions (MPI) and Underlying Events(UE)}

The hard scattering, parton shower, hadronisation, and secondary decays describe the primary final state arising from a high-energy parton–parton collision. However, since protons are colourless bound states of many coloured partons, the remnants of the protons—those partons which are not involved in the hard scatter—also influence the event. In the laboratory frame, high-energy protons appear as Lorentz-contracted, flattened discs that pass through each other during collision. The internal dynamics of these discs are effectively frozen during the extremely brief overlap, localizing the primary interaction but allowing other partons within the protons to interact independently \cite{MPIsummary,CorkeMPI}. This leads to the concept of multiple parton interactions (MPI), where several parton–parton scatterings occur within the same proton–proton collision. 

Perturbative QCD calculations, together with measured parton distribution functions (PDFs), predict that at high energies and low transverse momentum thresholds, the inclusive parton–parton cross section exceeds the total proton–proton cross section. This implies the presence of multiple semi-hard or hard scatterings per collision. Modelling MPI requires an additional non-perturbative input describing the spatial distribution of partons inside the proton, which governs the probability of multiple interactions. These additional scatterings, along with contributions from beam remnants, collectively form the underlying event—soft and semi-hard activity accompanying the hard process that significantly affects the observed final state \cite{mpi_general}.

\subsection{Diffraction and forward physics}
Diffractive processes characterized by rapidity gaps and (often) a leading proton are modelled in generators using Regge-inspired pictures (e.g., the Ingelman–Schlein Pomeron model \cite{IngelmanSchlein}). Practical treatments in general-purpose generators typically model diffractive systems as pseudo-collisions (Pomeron--proton) with their own PDFs and allow interleaving with MPI/ISR/FSR; alternative implementations (and dedicated hadronic models such as QGSJET/Epos) treat diffraction via enhanced Pomeron diagrams. Given the strong interplay between diffraction, MPI and gap survival effects, explicit references and tuned parameter sets are essential when comparing generators to forward/diffractive measurements \cite{SchulerSjostrand,Ostapchenko}.

\subsection{Hadronisation}

While the preceding stages of event generation rely on perturbative Quantum Chromodynamics (QCD), the observed final-state particles are colourless hadrons rather than free partons, as coloured partons cannot propagate freely due to confinement. Hadronization is the non-perturbative process by which coloured quarks and gluons transform into colour-neutral hadrons. Since first-principles calculations of hadronization from QCD are currently not possible, phenomenological models are employed to describe this stage.

Experimental observations from electron-positron annihilation into two-jet events show that hadrons are produced with characteristic distributions in rapidity and transverse momentum relative to the jet axis. The rapidity distribution is approximately flat within a certain range before falling off sharply, while the transverse momentum distribution is roughly Gaussian with a narrow width around 1–2 GeV, indicating most hadrons have low transverse momentum \cite{hard_processes}.

Early hadronization models, such as Independent Fragmentation Models, describe hadron production through iterative parton-to-hadron splittings guided by fragmentation functions fit to experimental data. However, these models have limitations, including frame dependence and a lack of a connection to QCD confinement dynamics. Modern hadronization models incorporate this picture of colour confinement by modelling the breaking of colour strings or clusters, converting the energy stored in the colour field into new hadrons, thereby producing the final observable particles \cite{hadronization}.

\subsection{Matching and Merging with Fixed-Order Calculations}

Fixed-order perturbative QCD calculations provide precise predictions for hard scattering processes involving a fixed number of partons, but they do not account for the all-order resummation of soft and collinear emissions. Parton showers, on the other hand, approximate this resummation but lack exact fixed-order accuracy for hard emissions. To combine the strengths of both approaches, matching and merging techniques have been developed. 
Matching ensures that the parton shower correctly reproduces the fixed-order matrix element for the lowest multiplicity process, avoiding double counting of emissions and preserving the accuracy of hard scattering predictions. Merging extends this concept to include multiple matrix elements with varying parton multiplicities, combining them consistently with the parton shower to provide an accurate description of both hard, well-separated jets and soft/collinear radiation. These techniques improve the reliability of event simulations across a wide range of energy scales and final-state topologies, and are essential for precise comparisons with collider data~\cite{matching_merging}.


\section{Event Generation in \textsc{Pythia}}
\label{pythia_evtgen}

\textsc{Pythia}~\cite{pythia_manual} is a widely used general-purpose event generator, implementing models for hard and soft interactions, multiparton interactions (MPIs), parton showers, hadronization, and decays. While Section~2 describes the general framework of Monte Carlo event generation, \textsc{Pythia} provides specific implementations and phenomenological models that distinguish it from other generators:  

\begin{itemize}
    \item \textbf{Parton showers:} Initial- and final-state radiation is modeled through backward and forward evolution algorithms, respectively, based on DGLAP splitting functions. \textsc{Pythia} uses transverse-momentum-ordered evolution, allowing flexible matching with fixed-order matrix elements. QED emissions, such as $q \to q\gamma$ and $l \to l\gamma$, are also included.

    \item \textbf{Hadronization:} The Lund string fragmentation model~\cite{lund-string} is employed. Colour confinement is modeled by stretching a string between outgoing partons; when the potential energy becomes sufficient, the string breaks, producing new quark-antiquark pairs iteratively until all partons form colour-singlet hadrons.

    \item \textbf{Multiparton interactions (MPI) and underlying event:} Multiple parton-parton interactions are simulated per event, with impact-parameter dependence controlling the probability of additional scatterings. A color-reconnection scheme improves the modeling of final-state hadronization. MPI modeling significantly contributes to the underlying event, especially in high-energy $pp$ collisions at the LHC.
\end{itemize}

This generator is optimized for general-purpose collider simulations and provides detailed phenomenological tuning parameters to reproduce experimental data across a wide range of observables.

\section{Event Generation in \textsc{Herwig}}
\label{herwig_evtgen}

\textsc{Herwig}~\cite{Herwig_2021} is another general-purpose event generator, distinguished by its coherent parton-shower algorithms and cluster hadronization approach. Building upon the framework outlined in Section~2, \textsc{Herwig} implements the following generator-specific features:  

\begin{itemize}
    \item \textbf{Parton showers:} Initial-state showers use a backward evolution algorithm with angular ordering, incorporating coherence effects and constraints on the emission angle. Final-state showers employ a coherent branching scheme where successive emissions respect angular ordering and energy sharing according to DGLAP splitting functions. Matrix-element corrections ensure accurate treatment of hard emissions.

    \item \textbf{Hadronization:} Using the pre-confinement property, outgoing partons are clustered into colour-singlet clusters with low invariant mass. These clusters then decay isotropically into hadrons. Heavier clusters may undergo iterative fission until all clusters are below a mass threshold, producing final-state hadrons. This cluster model contrasts with \textsc{Pythia}’s string fragmentation.

    \item \textbf{Multiparton interactions (MPI) and underlying event:} \textsc{Herwig} simulates MPI via additional secondary scatterings in events triggered by a single hard interaction. While the model reproduces average soft activity in the underlying event, it has limitations for minimum-bias events and diffractive contributions, especially in inclusive fiducial measurements at the LHC~\cite{MPI_HERWIG}.
\end{itemize}

\textsc{Herwig}’s focus on coherence and cluster-based hadronization provides an alternative approach to simulating QCD final states, complementary to \textsc{Pythia}’s string-based models.

\section{Comparison of models in \textsc{Pythia} and \textsc{Herwig} with Default settings at 7 and 13 TeV }
\label{comp_standalone}

The major difference between \textsc{Pythia} and \textsc{Herwig} lies primarily in the hadronization: the string model \cite{string_model,string_theory} in the first and the cluster model \cite{herwig-cluster,herwig_isr_fsr_had} in the second. 

Monte Carlo models are dependent upon various parameters that correspond to different physics phenomena. These parameters could be calibrated (tuned) to better describe the data. This section describes the comparison of  \textsc{Pythia} and \textsc{Herwig} using $pp$ collisions at $\sqrt{s}$ = 7~TeV, and 13~TeV, corresponding to Run 1 and Run 2, using a million events generated for each. The default input parameters used for the generation of events at $\sqrt{s}$ = 7~TeV and 13~TeV with \textsc{Pythia} and \textsc{Herwig} are listed in Tables~\ref{pythia_settings} and Table~\ref{herwig_settings}, respectively. In contrast, for the present study, we used \textsc{Herwig} 7.3, which incorporates significant developments in the parton shower, MPI, and hadronization models.  This choice was motivated by practical considerations: \textsc{Herwig} 7 is the actively supported version, and its framework is more convenient for direct comparison with \textsc{Pythia} in our analysis.

These parameter values correspond to the default settings as recommended by the \textsc{Pythia} \cite{pythia_manual} and \textsc{Herwig} authors \cite{herwig}, respectively.

\begin{table}[ht]
    \centering
    \begin{minipage}{0.485\textwidth}
        \centering
        \begin{tabular}{||c|c||}
            \hline
            Parameter & Value \\ \hline
            version & \textsc{Pythia}8.306 \\ \hline
            PDF & CT09MCS \\ \hline
            softQCD & on \\ \hline
            MPI & on \\ \hline
            alphaS & 0.130 \\ \hline
            pT0Ref & 2.28 \\ \hline
            $\epsilon$ & 0.215 \\ 
            \hline
        \end{tabular}
        \caption{\textsc{Pythia} parameters settings.}
        \label{pythia_settings}
    \end{minipage}\quad
    \begin{minipage}{0.485\textwidth}
        \centering
        \begin{tabular}{||c|c||}
            \hline
            Parameter & Value \\ \hline
            version & \textsc{Herwig}7.3  \\ \hline
            PDF & CT14nlo \\ \hline
            softQCD & on \\ \hline
            MPI & on \\ \hline 
            DLmode& 2 \\ \hline
            pTmin  & 3.1 \\ \hline
            $\epsilon$ & 0.21 \\
            \hline
            alphaS & 0.1185 \\
            \hline
        \end{tabular}
        \caption{\textsc{Herwig} parameters settings.}
        \label{herwig_settings}
    \end{minipage}
\end{table}

The PDF settings are crucial in event generators because they determine how the partons are distributed inside protons, which affects the simulated physics. CMS uses the CP5 tune with the NNPDF3.1 NNLO PDF set for \textsc{Pythia}. This tune provides a better description for underlying-event observables at LHC energies~\cite{cms_pythiatune,cms_tune}. ATLAS, on the other hand, uses its own dedicated tunes such as the A14 tune, which is based on the NNPDF2.3 LO PDF set~\cite{atlas_tune}. LHCb uses CT09MCS for \textsc{Pythia}, which is specifically designed for forward physics studies. "MCS" stands for modified charm strangeness, tailored to match LHCb's kinematic acceptance. For \textsc{Herwig}, CT14NLO provides a good balance between precision and consistency with theoretical calculations, making it suitable for LHCb’s Herwig-based simulations.

The DLmode parameter in \textsc{Herwig} describes the choice of Donnachie-Landshoff parametrization for the total cross section~\cite{dlmode}. The value set to 2 incorporates a two-pomeron model for a more refined treatment of diffractive scattering. The other parameters that control the MPI are $p^{T0}_{Ref}$, $\epsilon$ and $\alpha_{s}$. These parameters collectively govern the behavior of the system and shape the distributions of the physical quantities being studied. Eq.~\ref{factorisation} is divergent at low momentum transfers, and hence it could be regularized by the introduction of threshold parameter p$_{T0}$ as: 1/$p_{T}^{4}$ $\rightarrow$ 1/$(p_{T}^{2} + p_{T0}^{2})^{2}$. This approach supports the fact that at low $p_{T}$, the partons inside the proton are screened by one another, and at high values, the cross-section is small but nonzero. One also needs to include the center-of-mass energy dependency, since at higher energies partons are probed at smaller $x$, fraction of momentum carried away by the parton, where the parton density increases and the distance of the color screening decreases: 

\begin{equation}
p_{T0} = p_{T0}^{ref}\left(\frac{\sqrt{s}}{\sqrt{s_{0}}}\right)^\epsilon.
\label{pt0}
\end{equation}

The parameter $\sqrt{s_{0}}$ is given at a reference energy, $p_{T0}^{ref}$ is $p_{T0}$ at $\sqrt{s_{0}}$, $p_{T0}$ is a parameter that acts like a barrier between the hard and soft interactions, $\epsilon$ which defines the energy rescaling pace which needs to be tuned to experimental data. On the other hand, the strong coupling constant $\alpha_{s}$  also becomes divergent at low momentum and has to be regulated by a cut-off parameter.

Earlier versions of Herwig++ (2.7) used the UE‑EE‑5 tune~\cite{ue-ee-5}. However, ATLAS and CMS now use dedicated Herwig7 tunes such as CMS’s NNPDF3.1‑based sets and ATLAS’s H7-UE-MMHT tune, which are tailored to underlying event and multiple-parton interaction data in the context of \textsc{Herwig} 7~\cite{atlas_herwig7_tunes}.

The plots are generated for 1 million events in $pp$ collisions for the default configurations at $\sqrt{s}$ = 7 TeV and 13 TeV. The average number of particles produced in an event for both the default event generators at $\sqrt{s}$ = 7 and 13 TeV is shown in Tables~\ref{average_default_hadrons7} and \ref{average_default_hadrons13}, with the statistical uncertainties, respectively. At $\sqrt{s}$ = 13 TeV (Table~\ref{average_default_hadrons13}), the agreement between \textsc{Pythia} and \textsc{Herwig} is notably better than at $\sqrt{s}$ = 7 TeV, both in the total charged multiplicity and in the relative fractions of pions, kaons, and protons.

\begin{table*}[ht]
\small
\centering
\begin{tabular}{| *{9}{c|} }
    \hline
    & \multicolumn{2}{c|}{$N_{\pi}$}
    & \multicolumn{2}{c|}{$N_{K}$}
    & \multicolumn{2}{c|}{$N_{p}$}
    & \multicolumn{1}{c|}{$N_{ch}$} \\
    \hline
\textsc{Pythia} Default & 63.35 $\pm$ 0.07 & 77.1\%  
& 8.51 $\pm$ 0.01  & 10.3\% 
& 10.21 $\pm$ 0.01 & 12.4\% 
& 82.08 $\pm$ 0.09 \\
    \hline
\textsc{Herwig} Default & 54.78 $\pm$ 0.04 & 81.9\%
& 6.80 $\pm$ 0.01 & 10.17\%  
& 5.27 $\pm$ 0.01 & 7.86\% 
& 66.86 $\pm$ 0.06 \\
    \hline
\end{tabular}
\caption{Average number of charged hadrons in an event for the default settings of \textsc{Pythia} and \textsc{Herwig} at $\sqrt{s}$ = 7 TeV. The percentage indicates the fraction relative to the total average number of charged hadrons $N_{ch}$.}
\label{average_default_hadrons7}
\end{table*}

\begin{table*}[ht]
\small
\centering
\begin{tabular}{| *{9}{c|} }
    \hline
    & \multicolumn{2}{c|}{$N_{\pi}$}
    & \multicolumn{2}{c|}{$N_{K}$}
    & \multicolumn{2}{c|}{$N_{p}$}
    & \multicolumn{1}{c|}{$N_{ch}$} \\
    \hline
\textsc{Pythia} Default & 71.76 $\pm$ 0.08 & 77.5\%
& 9.69 $\pm$ 0.01  & 10.47\%
& 11.09 $\pm$ 0.01 & 11.98\%
& 92.55 $\pm$ 0.10 \\
\hline
\textsc{Herwig} Default & 75.13 $\pm$ 0.07 & 80.9\%
& 9.80 $\pm$ 0.01 & 10.55\%
& 7.89 $\pm$ 0.01 & 8.5\%
& 92.81 $\pm$ 0.08 \\
\hline
\end{tabular}
\caption{Average number of charged hadrons in an event for the default settings of \textsc{Pythia} and \textsc{Herwig} at $\sqrt{s} = 13$ TeV. The percentage indicates the fraction relative to the total average number of charged hadrons $N_{ch}$.}
\label{average_default_hadrons13}
\end{table*}


\begin{figure*}[!htbp]
  \centering

  \begin{subfigure}[b]{0.45\linewidth}
    \centering
    \includegraphics[width=\linewidth]{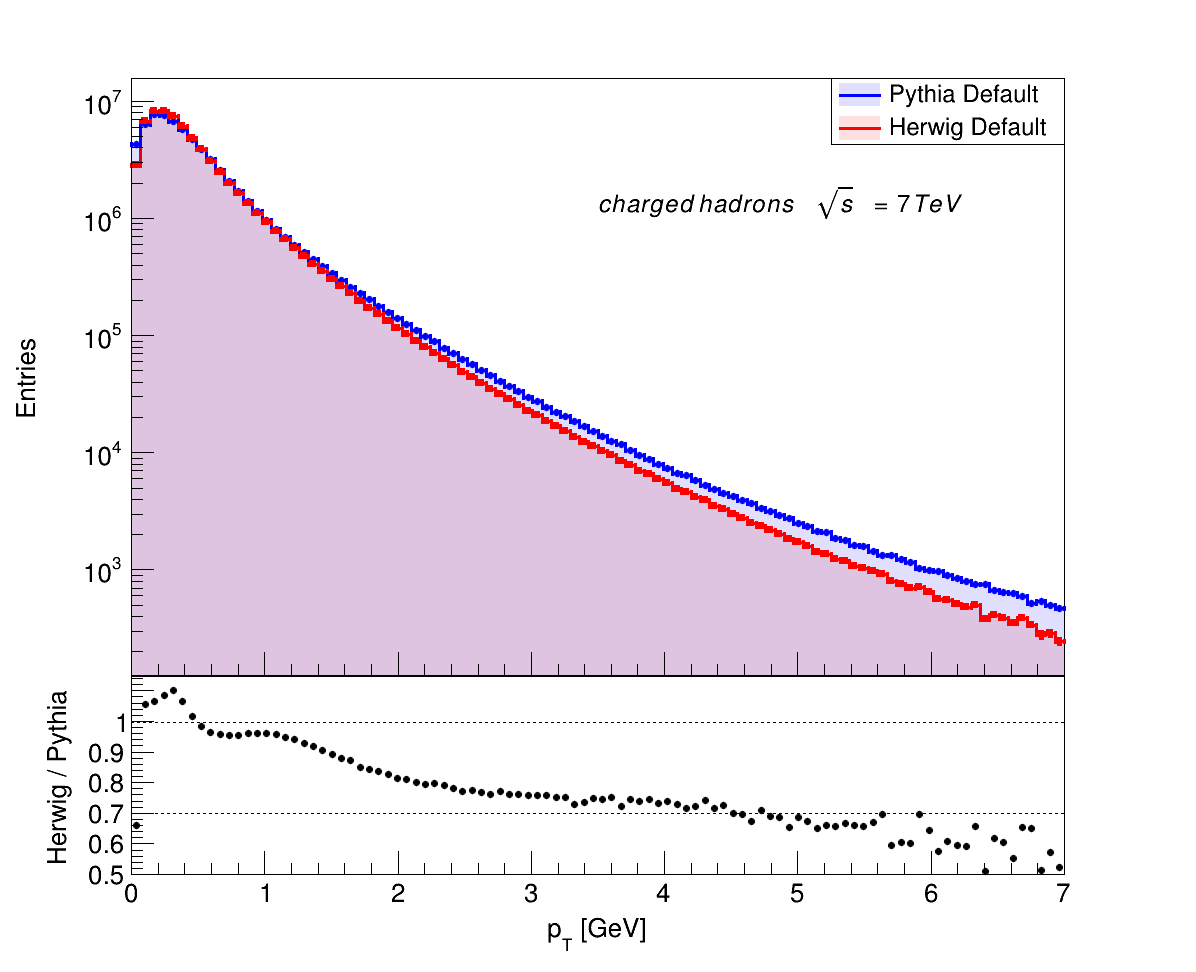}
    \caption{}
  \end{subfigure}
  \hspace{-1ex}
  \begin{subfigure}[b]{0.45\linewidth}
    \centering
    \includegraphics[width=\linewidth]{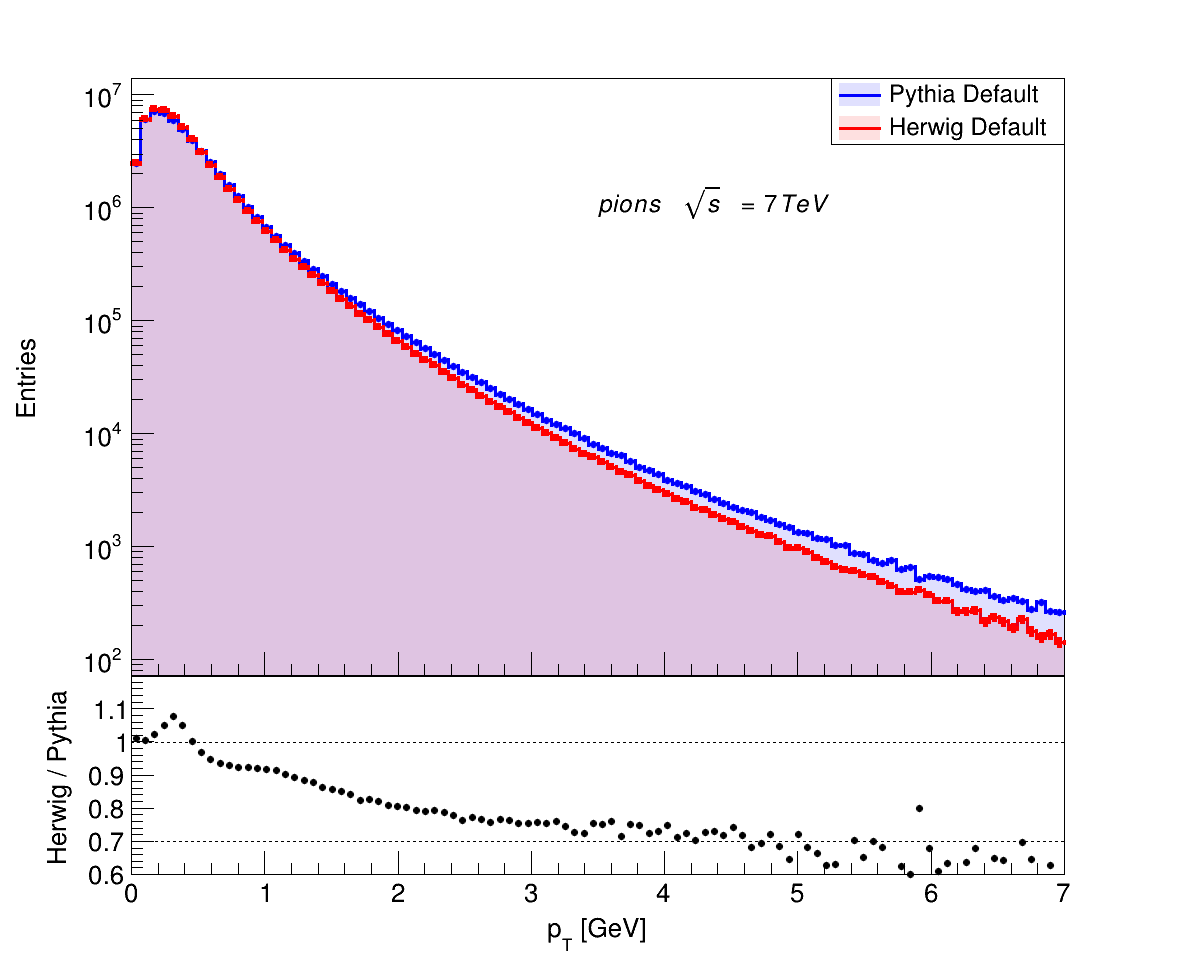}
    \caption{}
  \end{subfigure}

  \vspace{-1ex}

  \begin{subfigure}[b]{0.45\linewidth}
    \centering
    \includegraphics[width=\linewidth]{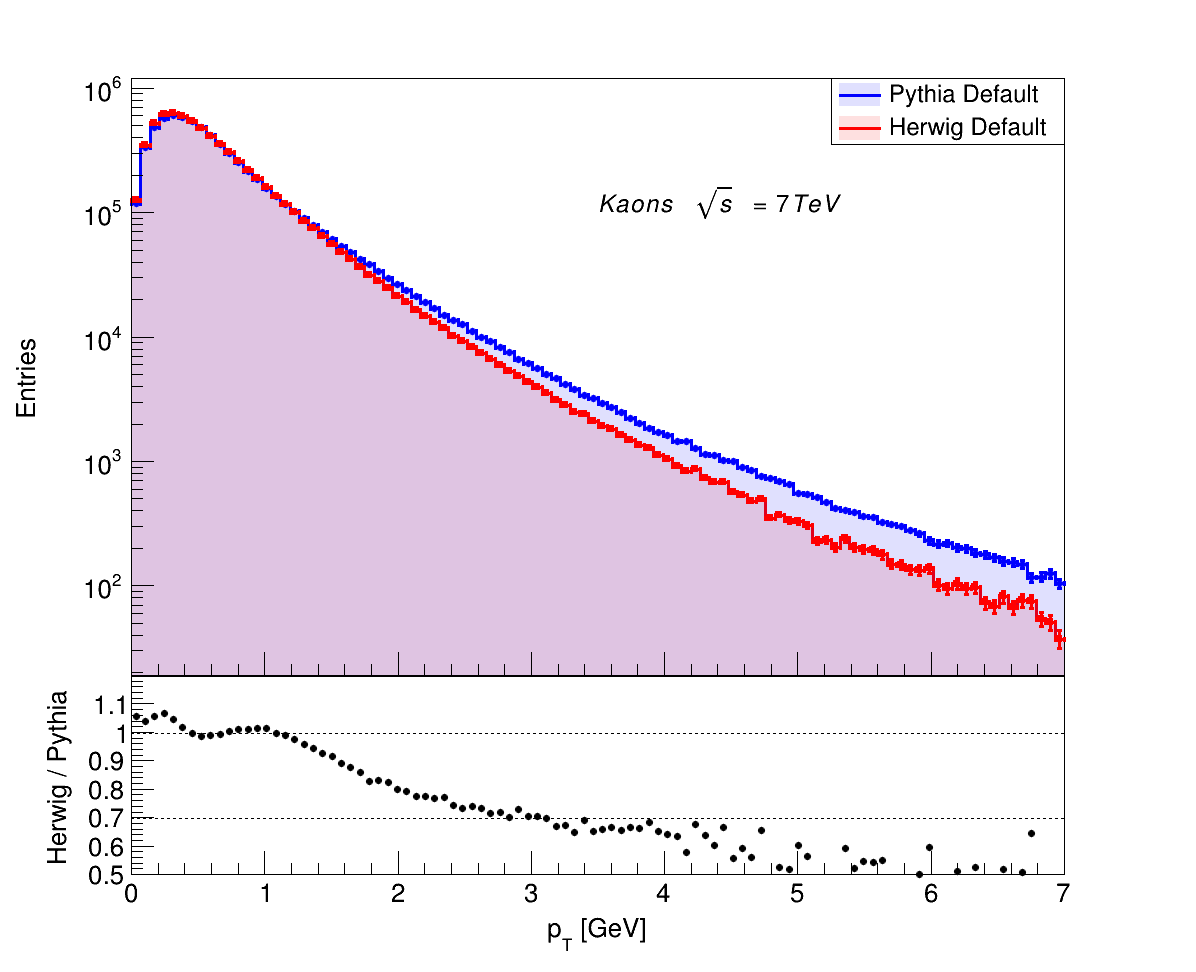}
    \caption{}
  \end{subfigure}
  \hspace{-1ex}
  \begin{subfigure}[b]{0.45\linewidth}
    \centering
    \includegraphics[width=\linewidth]{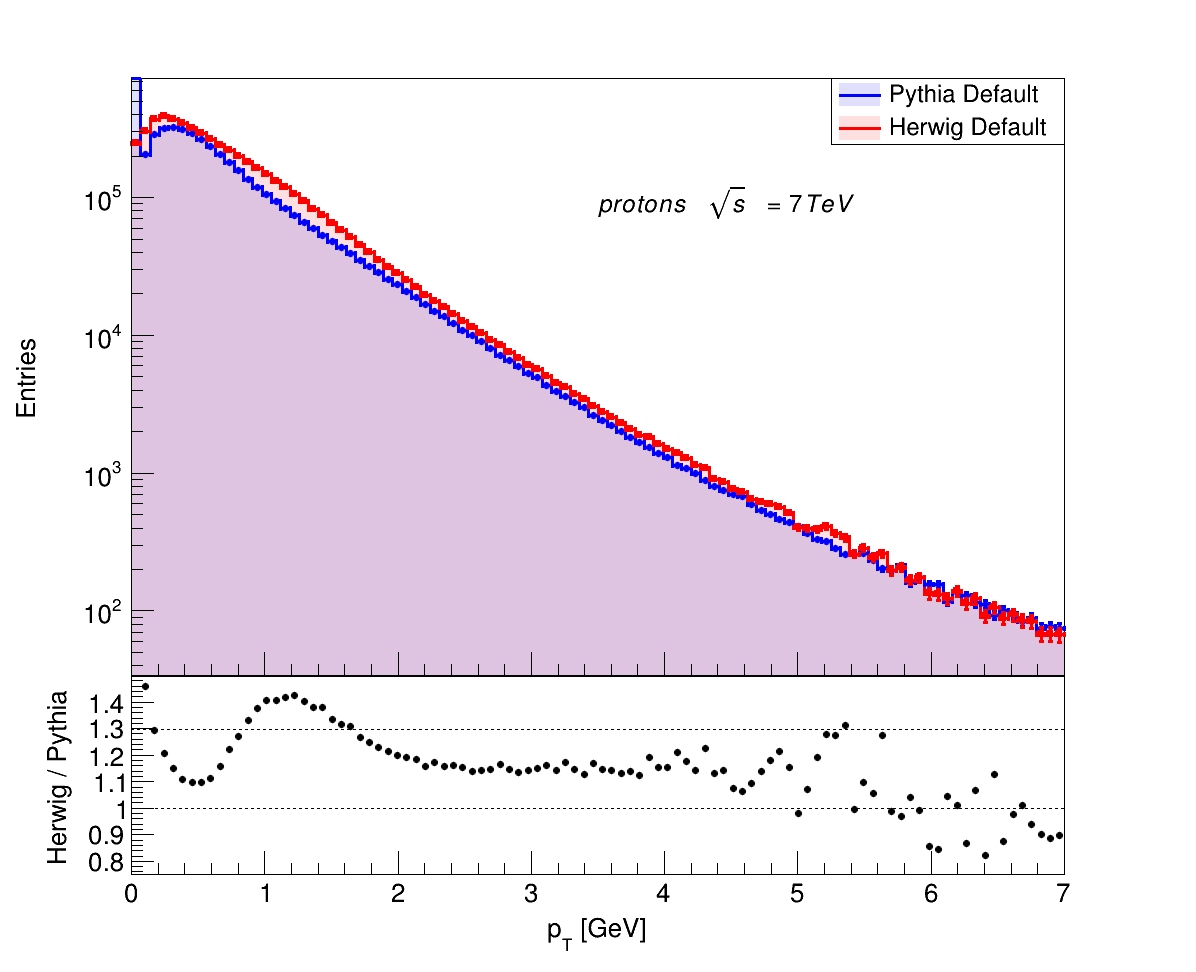}
    \caption{}
  \end{subfigure}

  \caption{Comparison of \textsc{Pythia} and \textsc{Herwig} transverse momentum distributions at $\sqrt{s} = 7~\mathrm{TeV}$ for (a) charged hadrons, (b) pions, (c) kaons, and (d) protons.}
  \label{pyth_herwig_comparison_pt7}
\end{figure*}

\begin{figure*}[!htbp]
  \centering

  \begin{subfigure}[b]{0.45\linewidth}
    \centering
    \includegraphics[width=\linewidth]{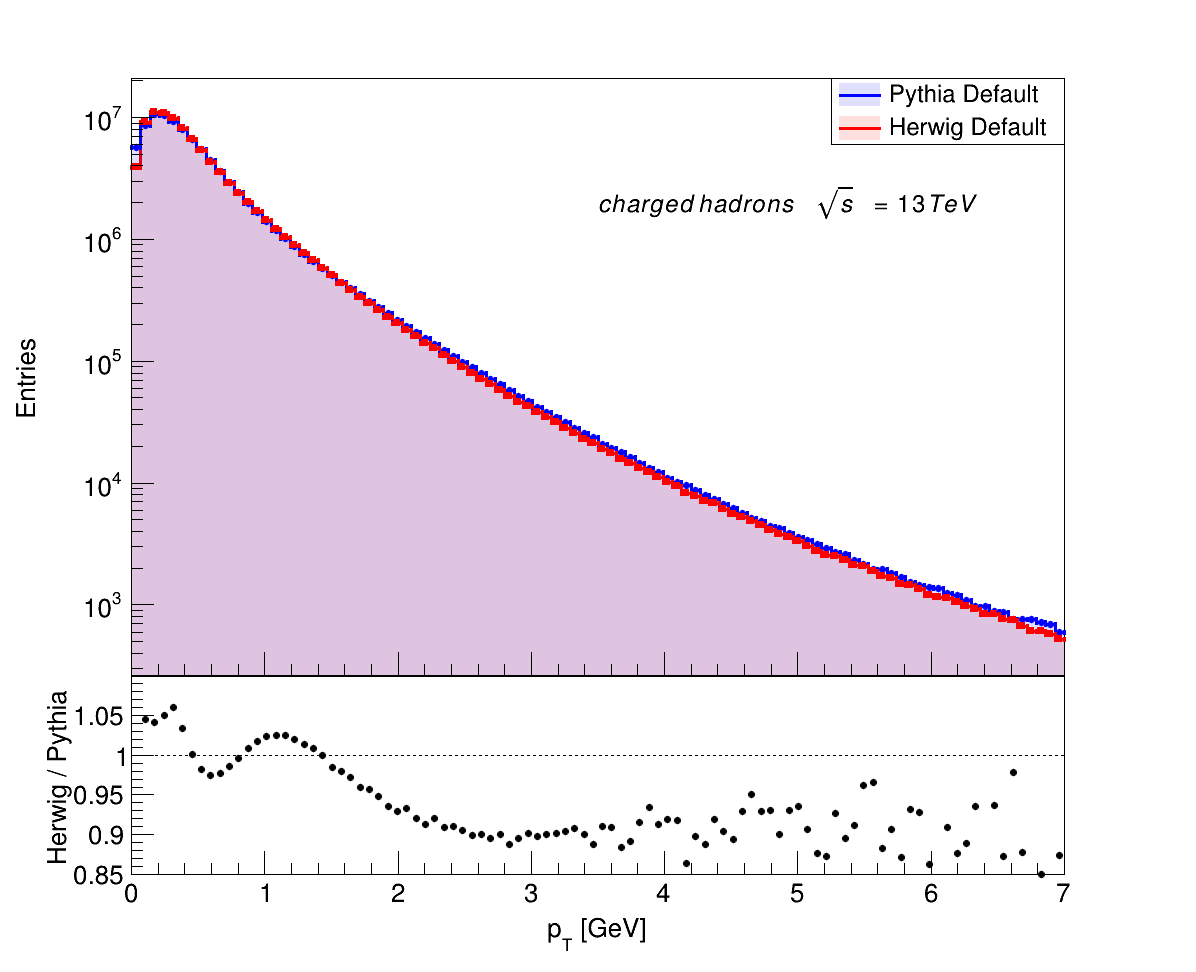}
    \caption{}
  \end{subfigure}
  \hspace{-1ex}
  \begin{subfigure}[b]{0.45\linewidth}
    \centering
    \includegraphics[width=\linewidth]{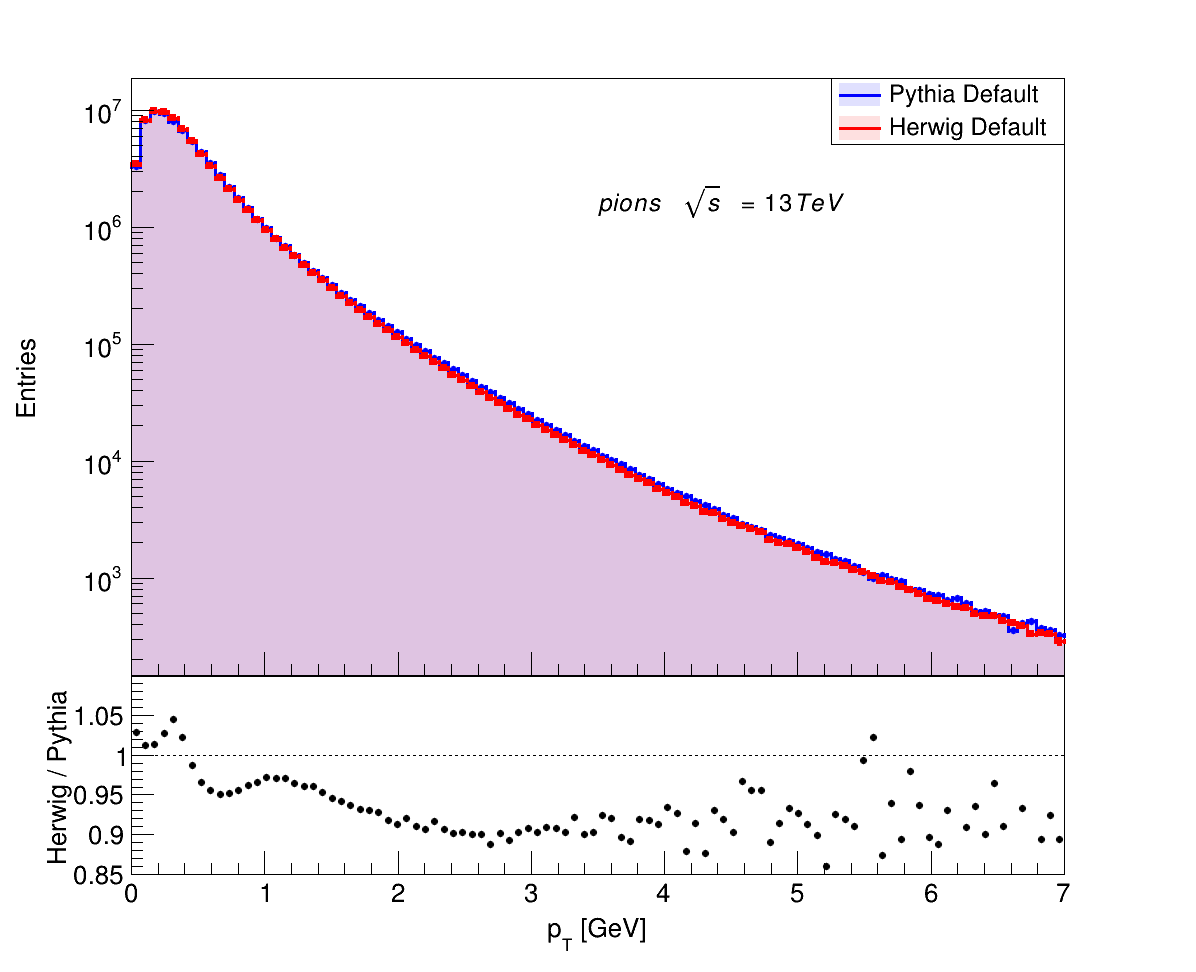}
    \caption{}
  \end{subfigure}

  \vspace{2ex}

  \begin{subfigure}[b]{0.45\linewidth}
    \centering
    \includegraphics[width=\linewidth]{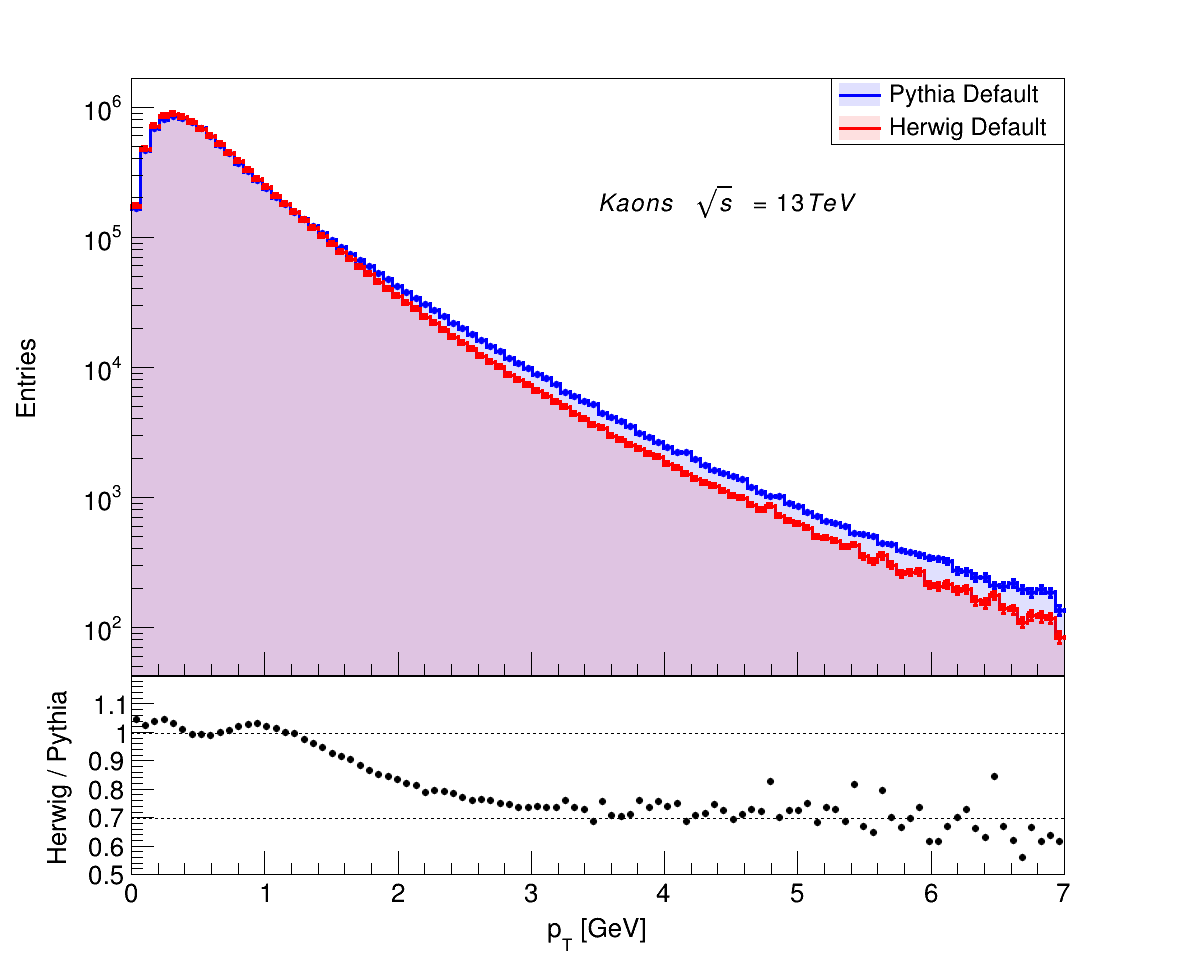}
    \caption{}
  \end{subfigure}
  \hspace{-1ex}
  \begin{subfigure}[b]{0.45\linewidth}
    \centering
    \includegraphics[width=\linewidth]{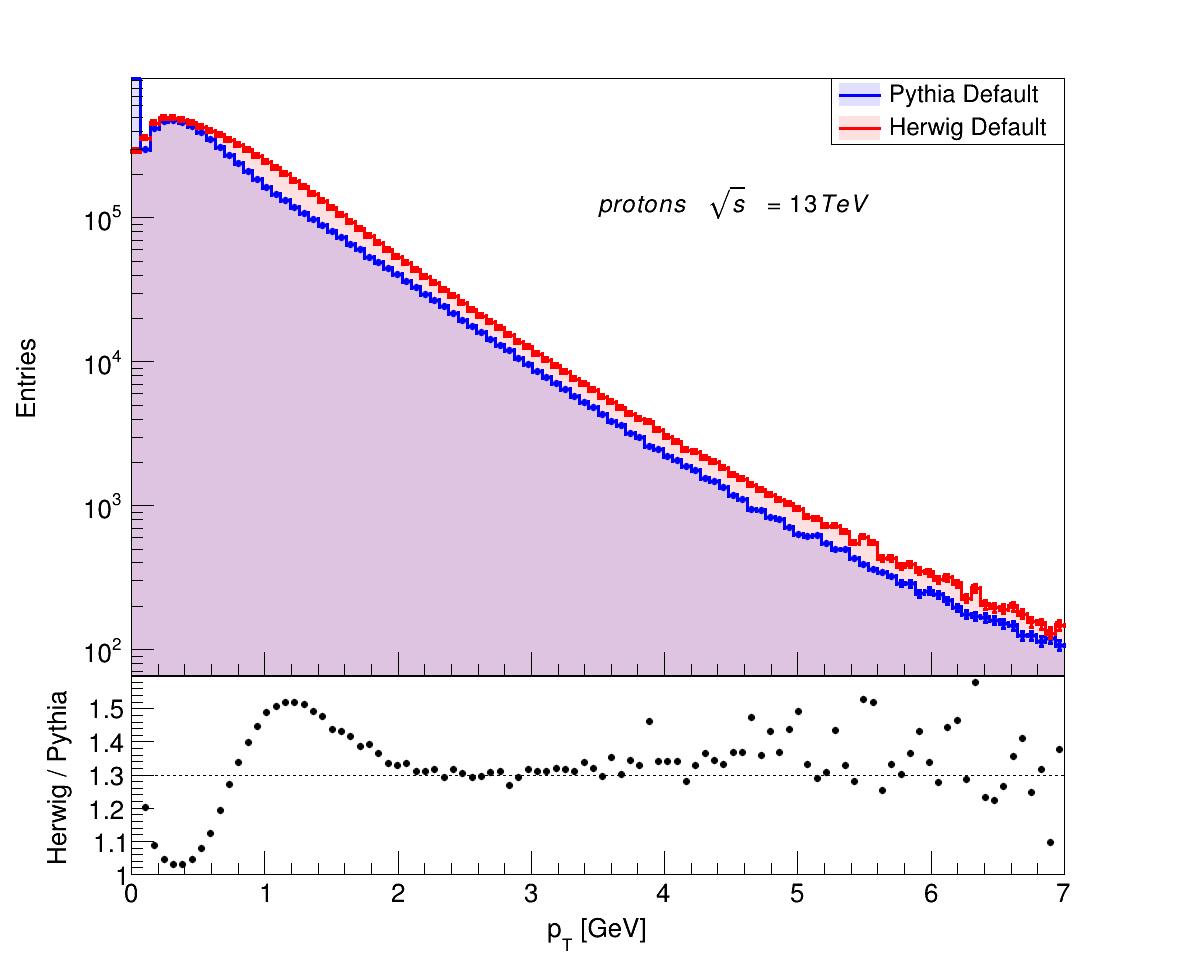}
    \caption{}
  \end{subfigure}

  \caption{Comparison of \textsc{Pythia} and \textsc{Herwig} transverse momentum distributions at $\sqrt{s} = 13~\mathrm{TeV}$ for (a) charged hadrons, (b) pions, (c) kaons, and (d) protons.}
  \label{pyth_herwig_comparison_pt13}
\end{figure*}

Figure~\ref{pyth_herwig_comparison_pt7} illustrates the distribution of transverse momentum of charged hadrons produced in $pp$ collisions at $\sqrt{s}$ = 7 TeV in (a), while (b), (c), and (d) represent the distribution of pions, kaons, and protons, respectively. The error bars represent statistical uncertainties.  The ratio of \textsc{Pythia} to \textsc{Herwig} is shown in the lower panel of each plot, indicating that the ratio subceeds unity at higher momentum scales for the charged hadrons, which is quite evident also in the pions and kaons plots showing that \textsc{Pythia} produces more particles at higher momentum values as compared to \textsc{Herwig}. The case is opposite for protons; the distribution is above unity, indicating that more low-momentum protons are produced by \textsc{Herwig} than \textsc{Pythia}. A similar trend is also observed at 13 TeV, as shown in ~Fig.~\ref{pyth_herwig_comparison_pt13}, although it is less pronounced compared to lower energies.


\begin{figure*}[!htbp]
  \centering

  \begin{subfigure}[b]{0.45\linewidth}
    \centering
    \includegraphics[width=\linewidth]{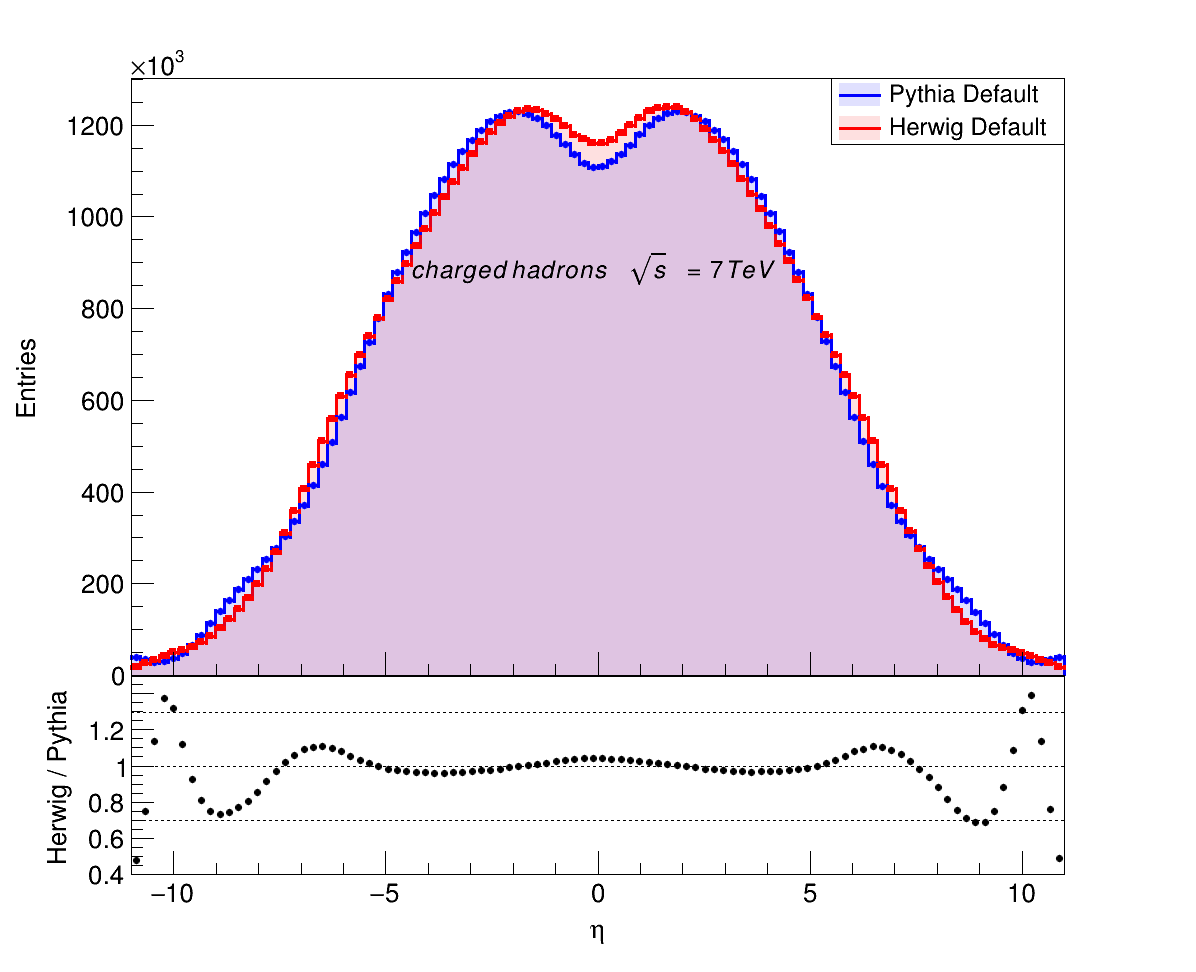}
    \caption{}
    \label{fig:eta_charged_7}
  \end{subfigure}
  \hspace{-1ex}
  \begin{subfigure}[b]{0.45\linewidth}
    \centering
    \includegraphics[width=\linewidth]{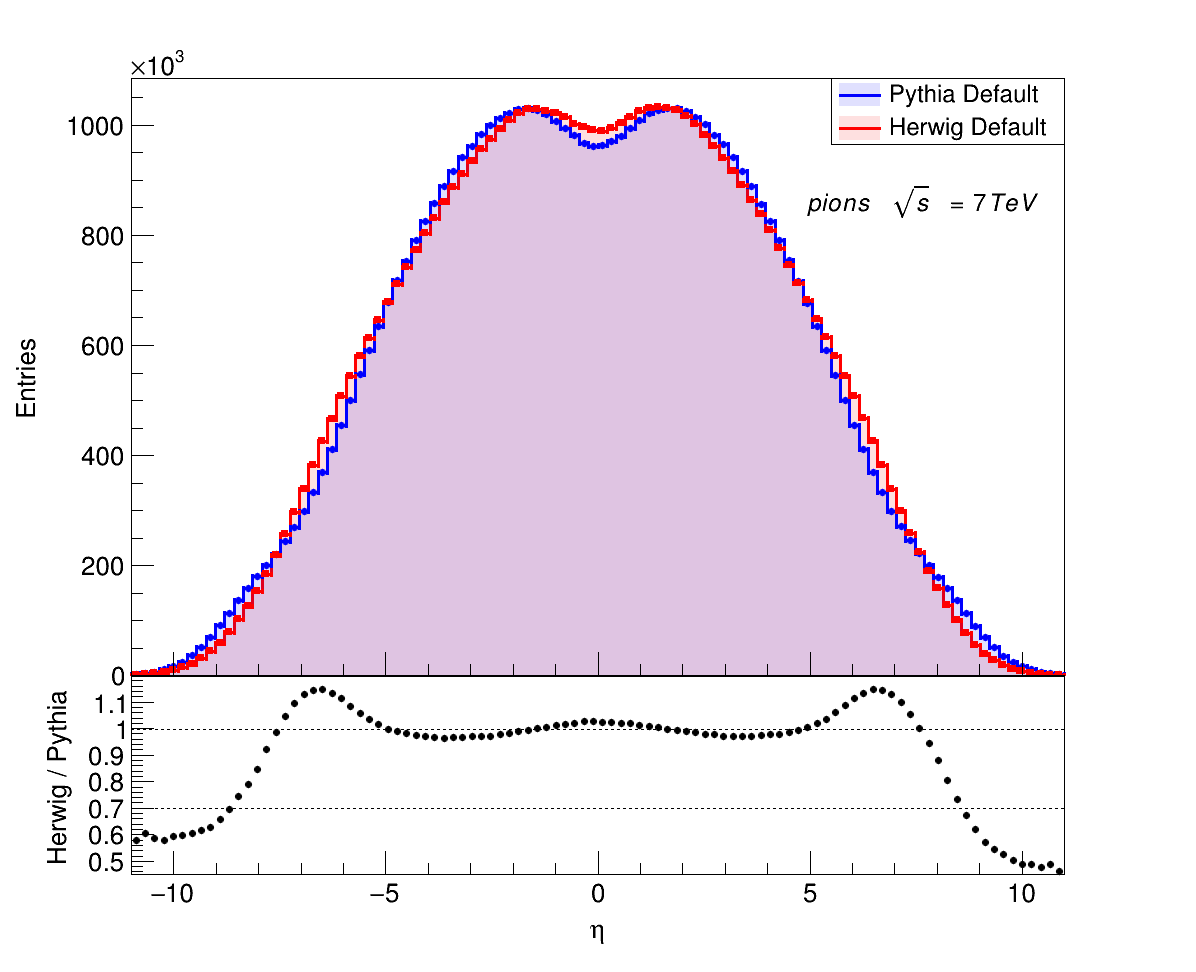}
    \caption{}
    \label{fig:eta_pions_7}
  \end{subfigure}

  \vspace{2ex}

  \begin{subfigure}[b]{0.45\linewidth}
    \centering
    \includegraphics[width=\linewidth]{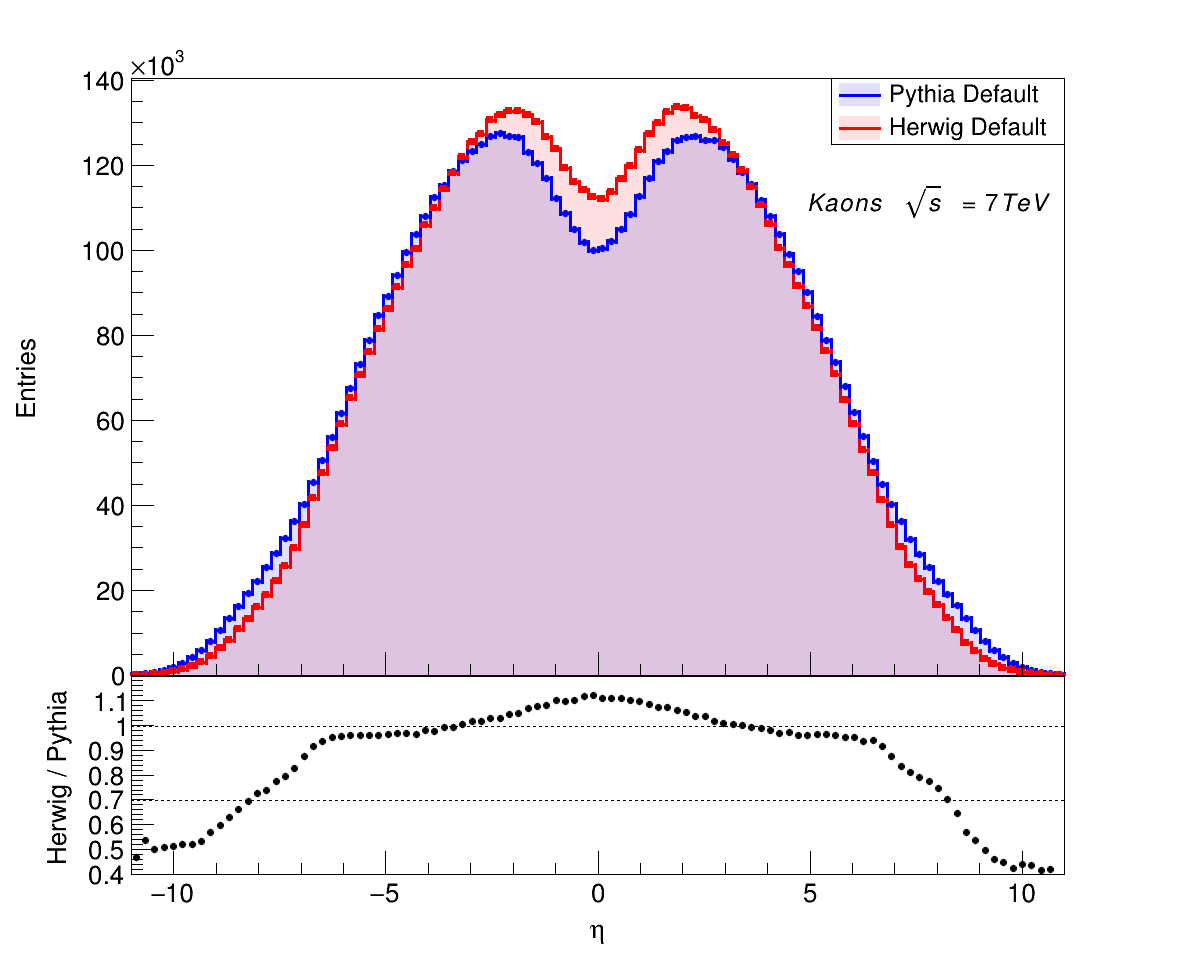}
    \caption{}
    \label{fig:eta_kaons_7}
  \end{subfigure}
  \hspace{-1ex}
  \begin{subfigure}[b]{0.45\linewidth}
    \centering
    \includegraphics[width=\linewidth]{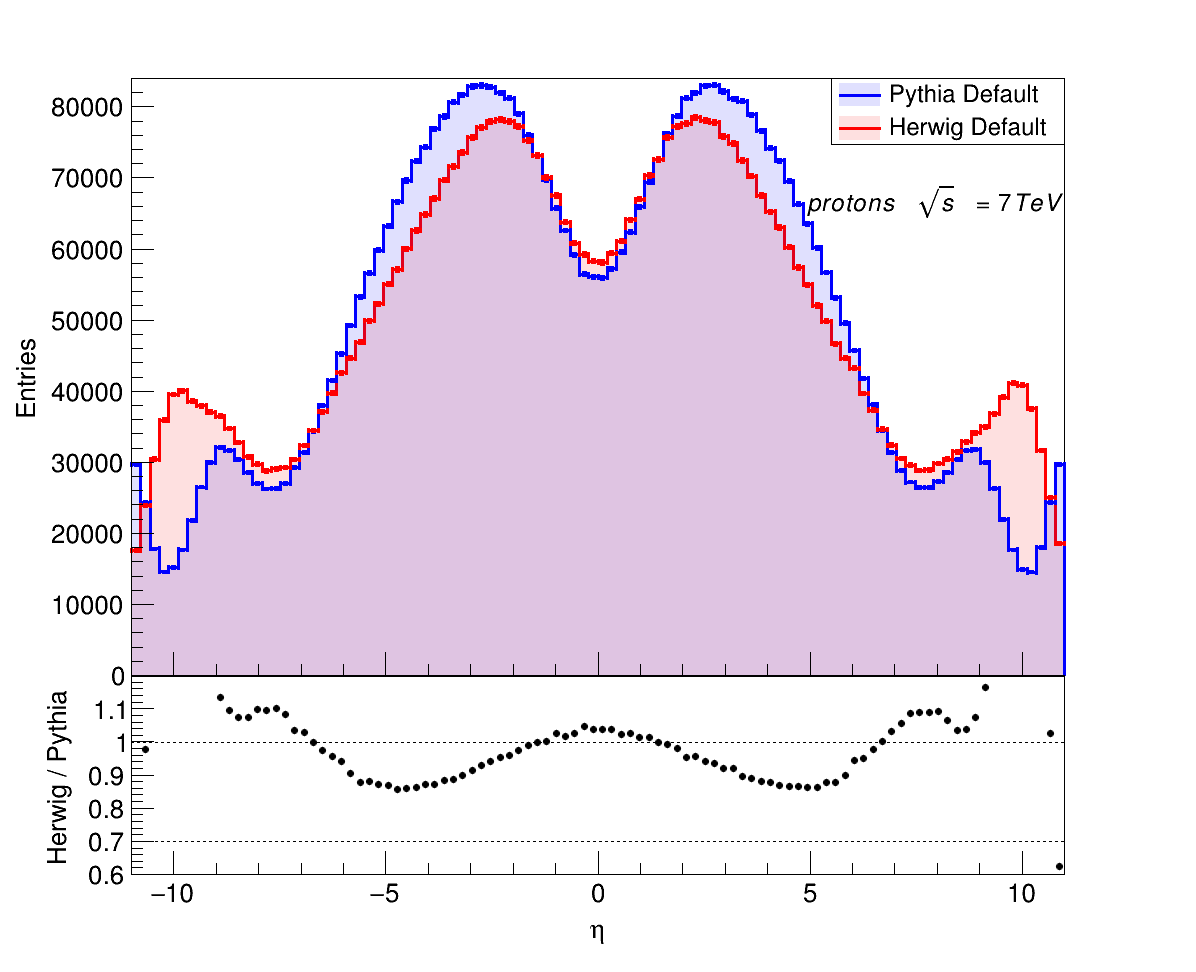}
    \caption{}
    \label{fig:eta_protons_7}
  \end{subfigure}

  \caption{Comparison of \textsc{Pythia} and \textsc{Herwig} pseudorapidity ($\eta$) distributions at $\sqrt{s} = 7~\mathrm{TeV}$ for (a) charged hadrons, (b) pions, (c) kaons, and (d) protons. The error bars represent statistical uncertainties.}
  \label{pyth_herwig_comparison_eta7}
\end{figure*}


\begin{figure*}[!htbp]
  \centering

  \begin{subfigure}[b]{0.45\linewidth}
    \centering
    \includegraphics[width=\linewidth]{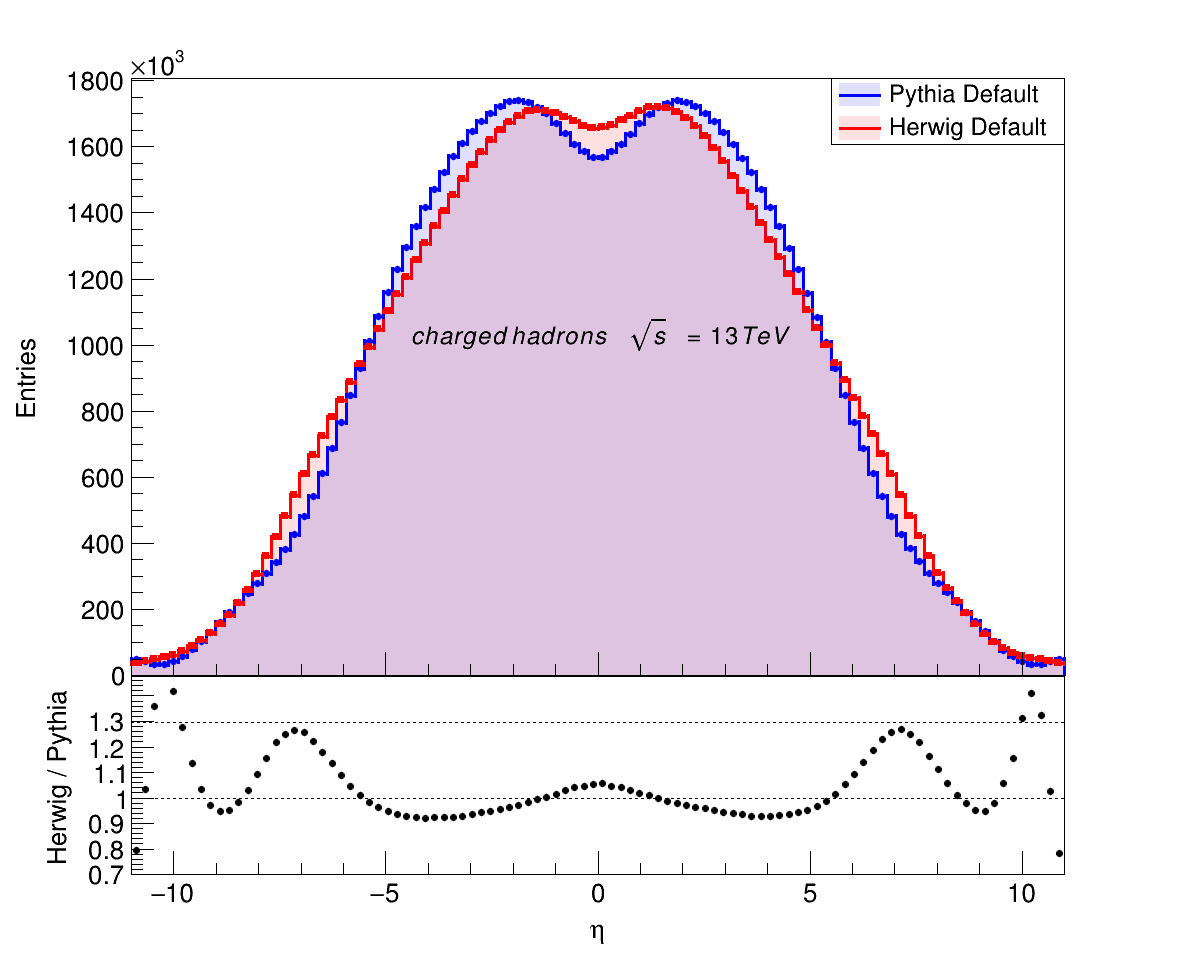}
    \caption{}
    \label{fig:eta_charged_13}
  \end{subfigure}
  \hspace{-1ex}
  \begin{subfigure}[b]{0.45\linewidth}
    \centering
    \includegraphics[width=\linewidth]{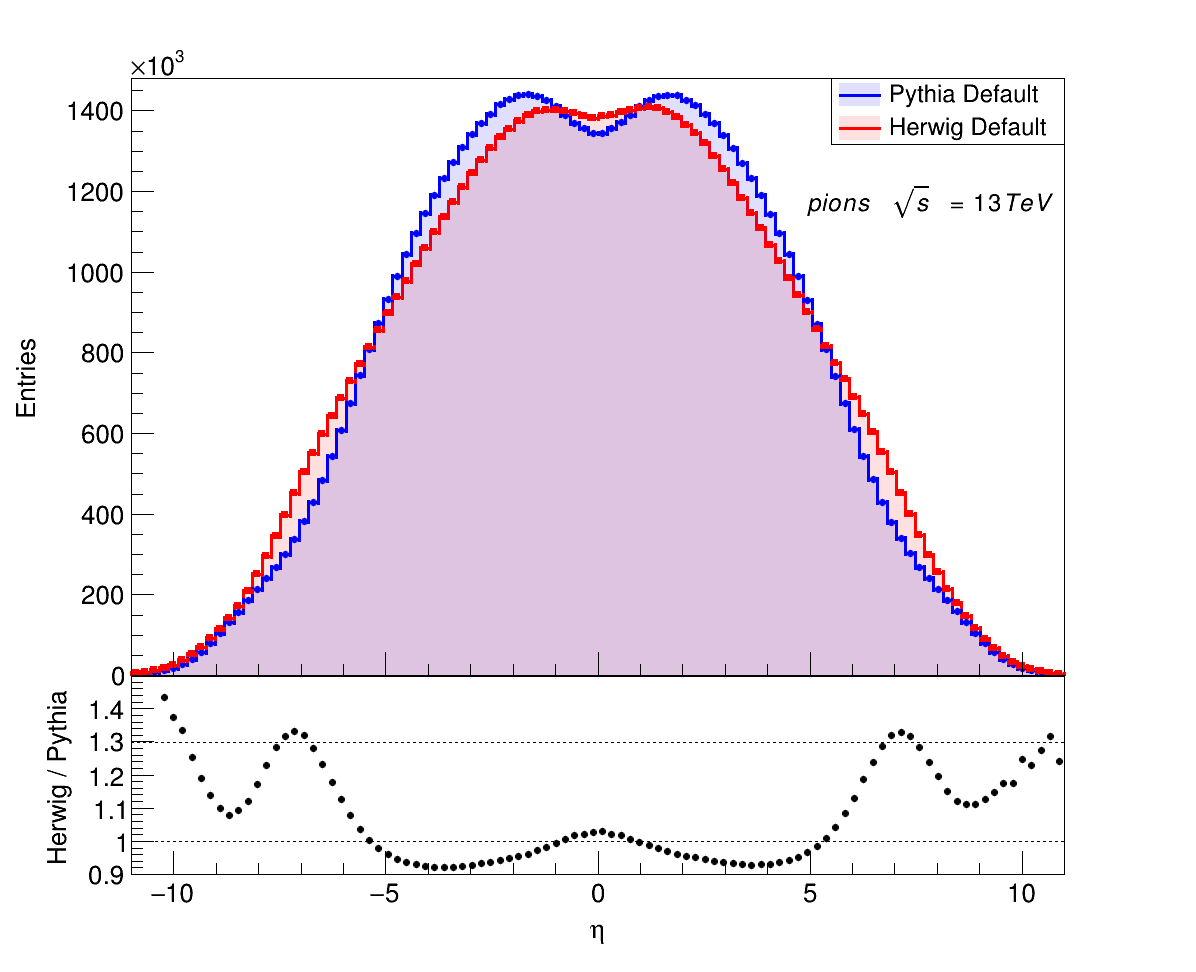}
    \caption{}
    \label{fig:eta_pions_13}
  \end{subfigure}

  \vspace{2ex}

  \begin{subfigure}[b]{0.45\linewidth}
    \centering
    \includegraphics[width=\linewidth]{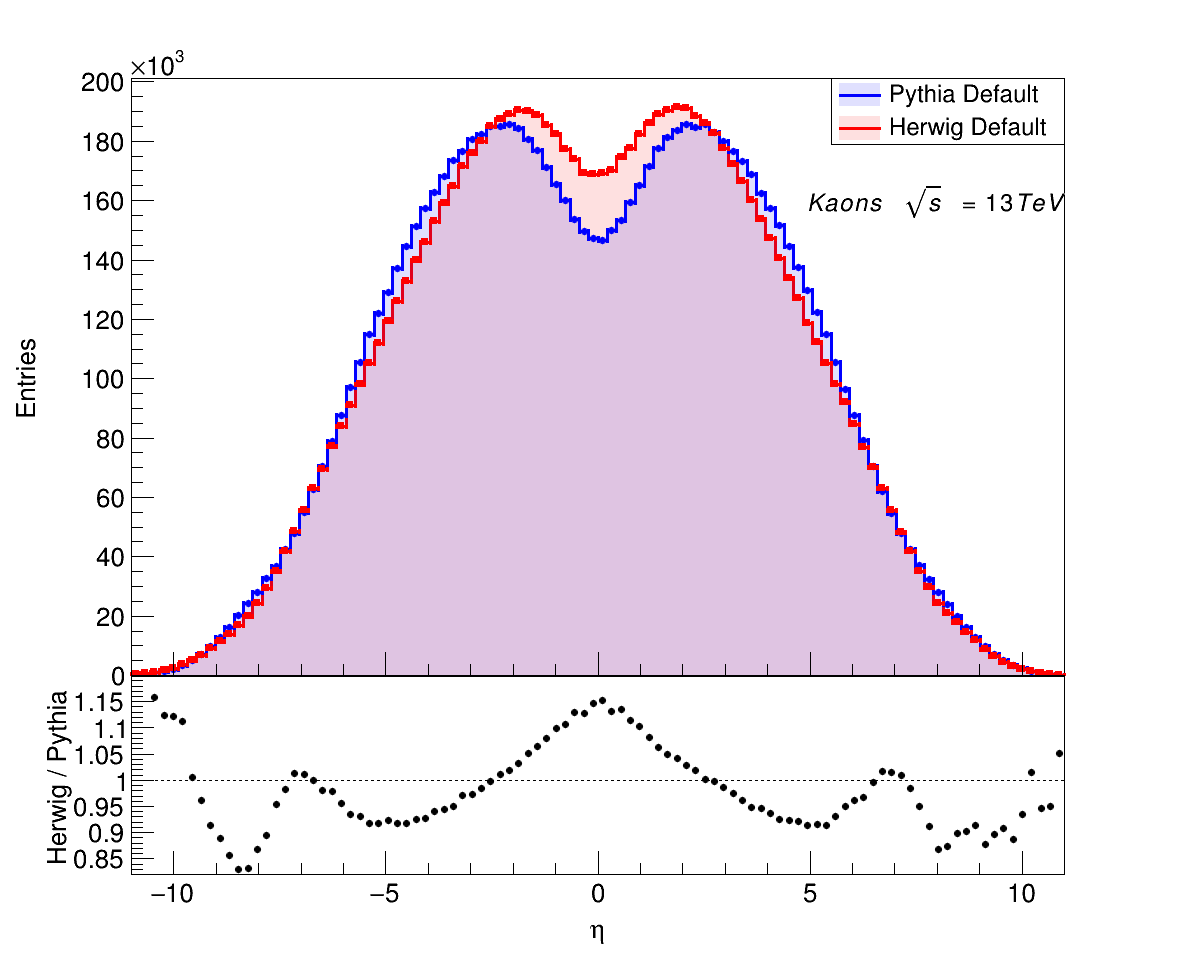}
    \caption{}
    \label{fig:eta_kaons_13}
  \end{subfigure}
  \hspace{-1ex}
  \begin{subfigure}[b]{0.45\linewidth}
    \centering
    \includegraphics[width=\linewidth]{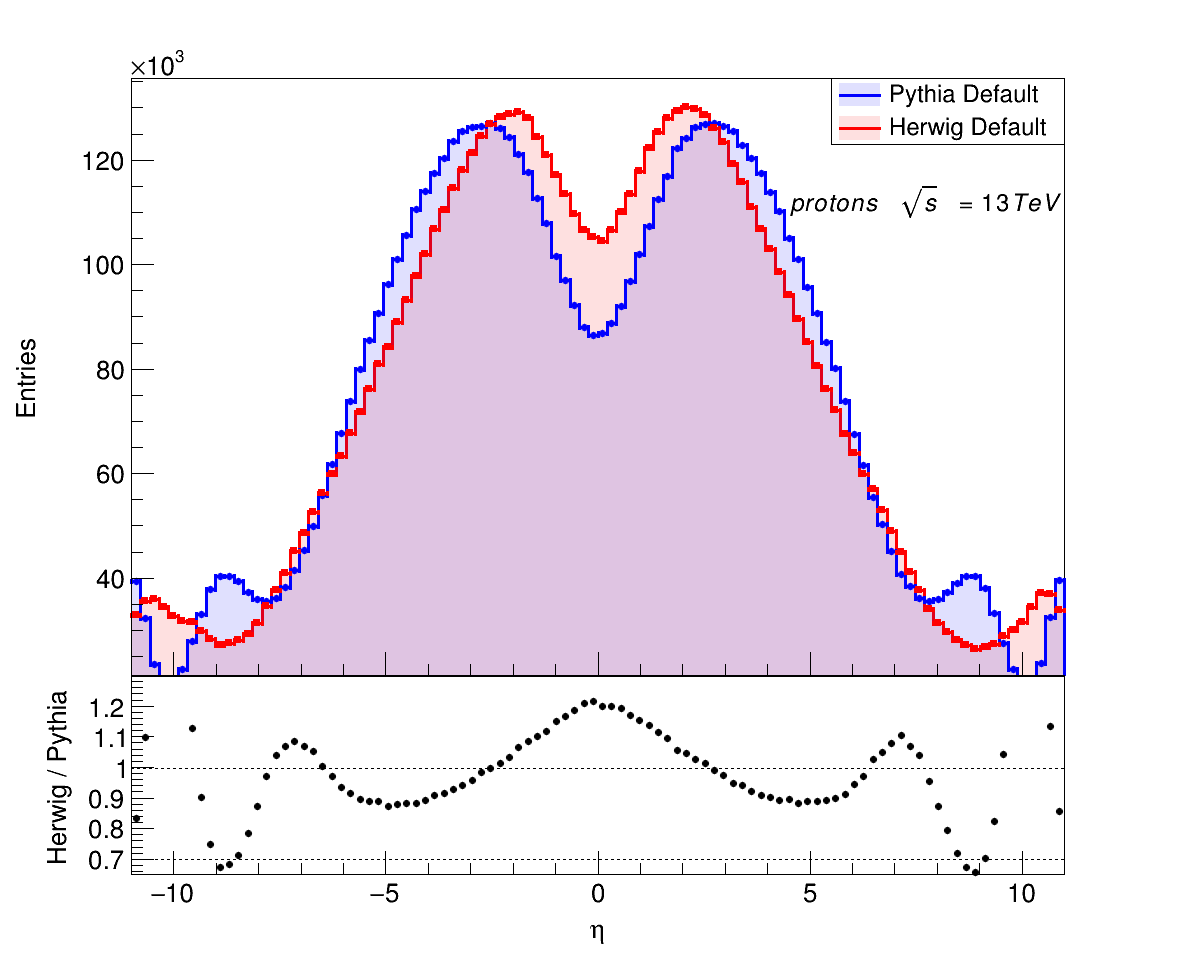}
    \caption{}
    \label{fig:eta_protons_13}
  \end{subfigure}

  \caption{Comparison of \textsc{Pythia} and \textsc{Herwig} pseudorapidity ($\eta$) distributions at $\sqrt{s} = 13~\mathrm{TeV}$ for (a) charged hadrons, (b) pions, (c) kaons, and (d) protons. The error bars represent statistical uncertainties.}
  \label{pyth_herwig_comparison_eta13}
\end{figure*}

Figure~\ref{pyth_herwig_comparison_eta7} shows the comparison of \textsc{Pythia} and \textsc{Herwig} at 7 TeV for the pseudorapidity distribution $\eta$, with statistical uncertainties, of charged hadrons (a), pions (b), kaons (c), and protons (d). \textsc{Pythia} and \textsc{Herwig} predictions show small differences in the forward regions of $\vert\eta\vert > 5$ and slight variations near the peak at $\eta$ = 0. These differences are likely due to the distinct underlying physics model in hadronization and parton showering, or to the tuning of model parameters. Since the majority of the particles in the charged distribution are pions, pions follow a similar trend. As could be seen from the ratio panel, there is a good general agreement for both by the two generators in the -3$ < \eta <$3 region. For the case of kaons, the central region is dominated by high-energy particle production from the hardest partonic interactions, where the models tend to disagree. Lastly, for the case of protons, there is a difference between the distribution for \textsc{Herwig} and \textsc{Pythia}, indicating more elastic protons in \textsc{Pythia} than \textsc{Herwig}. There are also some noticeable peaks at $|\eta| \sim$ 8 and 10, indicating that a significant fraction of protons are produced in the forward and backward regions. These protons are remnants of the incoming protons from the initial collision, which are typically associated with soft interactions, and the difference in the models is because of the models used in \textsc{Pythia} and \textsc{Herwig} for the treatment of MPI, diffraction, and proton remnants. A similar trend can be seen in the $\eta$ distributions at 13~TeV as well, which is shown in Fig.~\ref{pyth_herwig_comparison_eta13}. Both generators are in reasonable agreement in the central region, but diverge in the forward/backward regions.


\begin{figure*}[!htbp]
  \centering

  \begin{subfigure}[b]{0.45\linewidth}
    \centering
    \includegraphics[width=\linewidth]{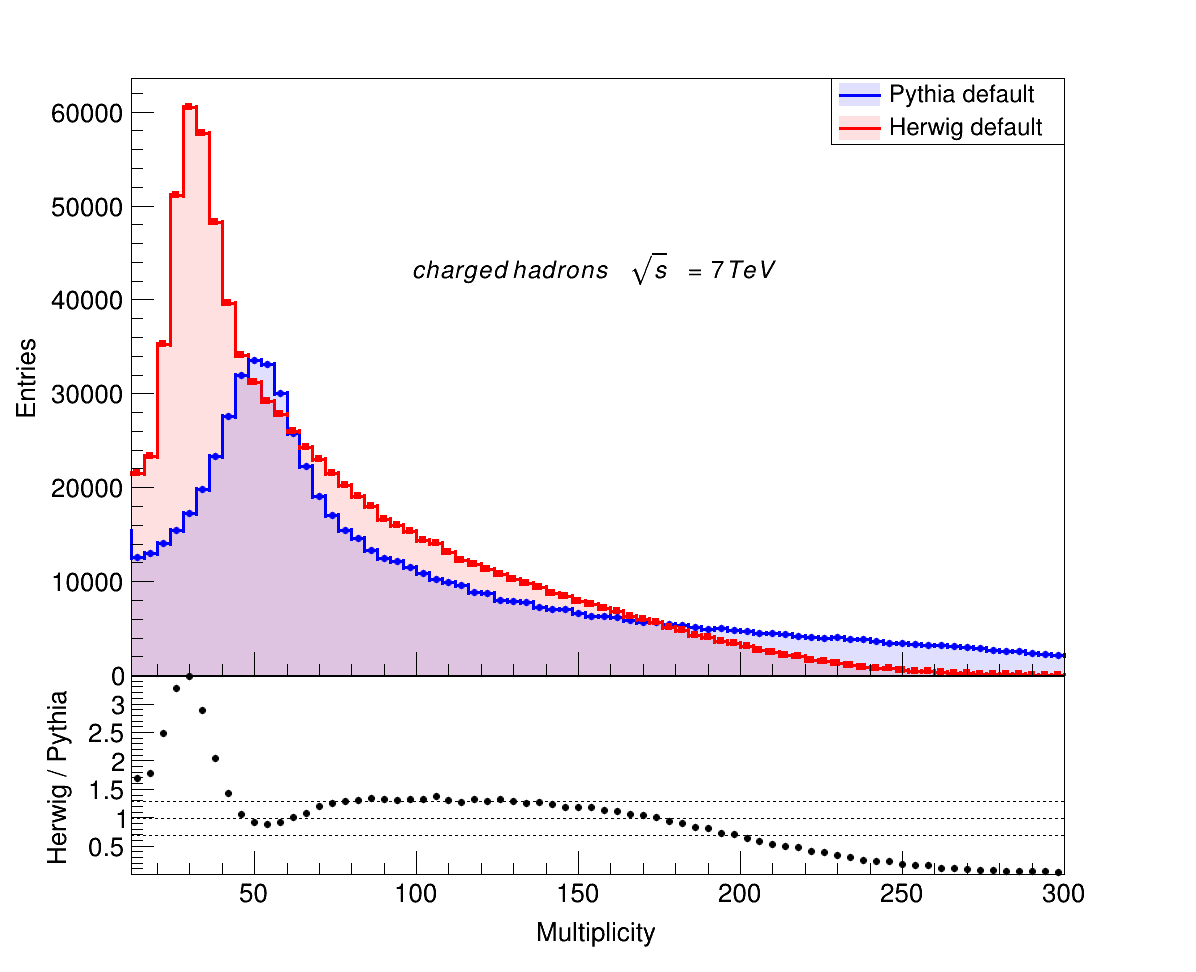}
    \caption{}
    \label{fig:mult_charged_7}
  \end{subfigure}
  \hspace{-1ex}
  \begin{subfigure}[b]{0.45\linewidth}
    \centering
    \includegraphics[width=\linewidth]{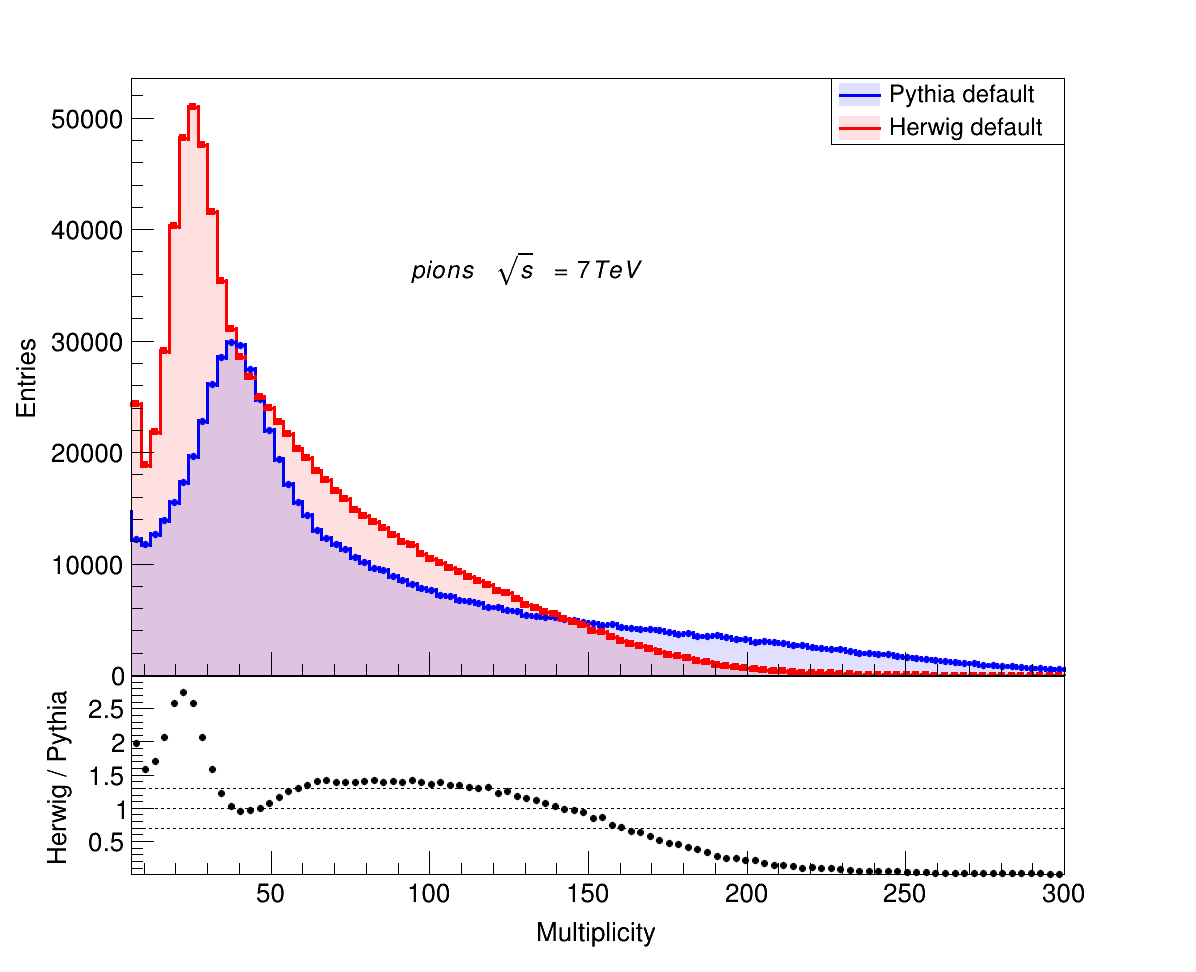}
    \caption{}
    \label{fig:mult_pions_7}
  \end{subfigure}

  \vspace{2ex}

  \begin{subfigure}[b]{0.45\linewidth}
    \centering
    \includegraphics[width=\linewidth]{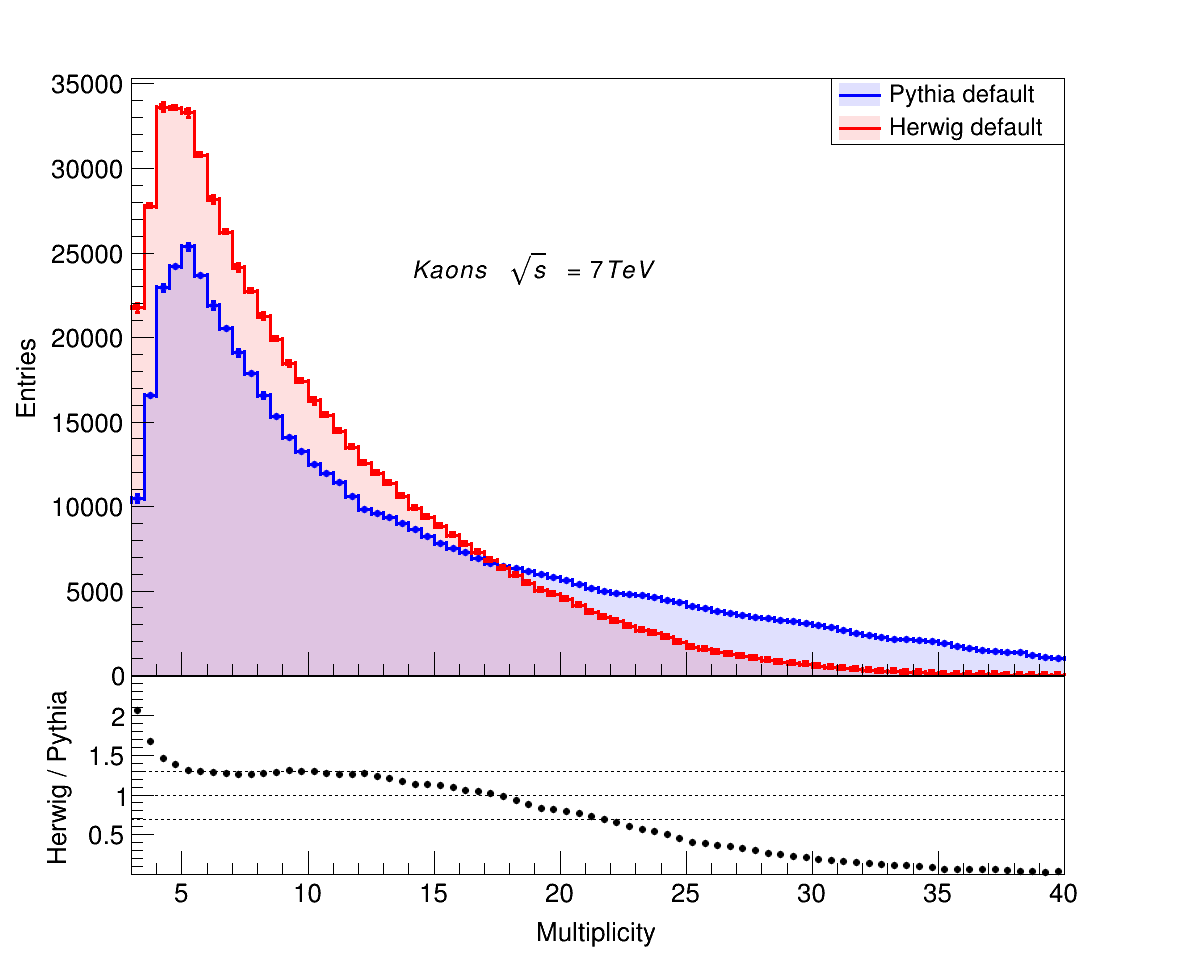}
    \caption{}
    \label{fig:mult_kaons_7}
  \end{subfigure}
  \hspace{-1ex}
  \begin{subfigure}[b]{0.45\linewidth}
    \centering
    \includegraphics[width=\linewidth]{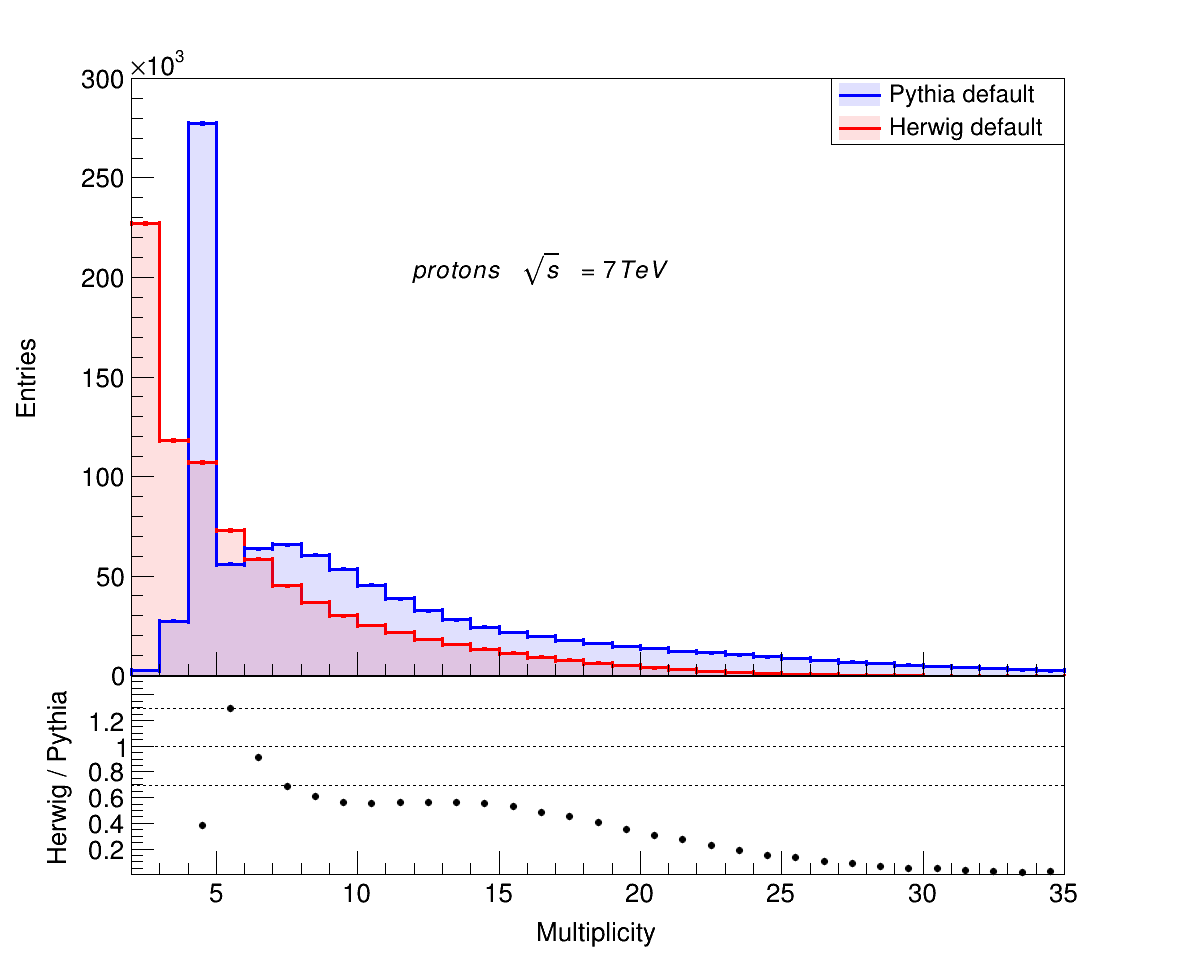}
    \caption{}
    \label{fig:mult_protons_7}
  \end{subfigure}

  \caption{Comparison of \textsc{Pythia} and \textsc{Herwig} multiplicity distributions at $\sqrt{s} = 7~\mathrm{TeV}$ for (a) charged hadrons, (b) pions, (c) kaons, and (d) protons. The error bars represent statistical uncertainties. }
  \label{pyth_herwig_comparison_mult7}
\end{figure*}

\begin{figure*}[!htbp]
  \centering

  \begin{subfigure}[b]{0.45\linewidth}
    \centering
    \includegraphics[width=\linewidth]{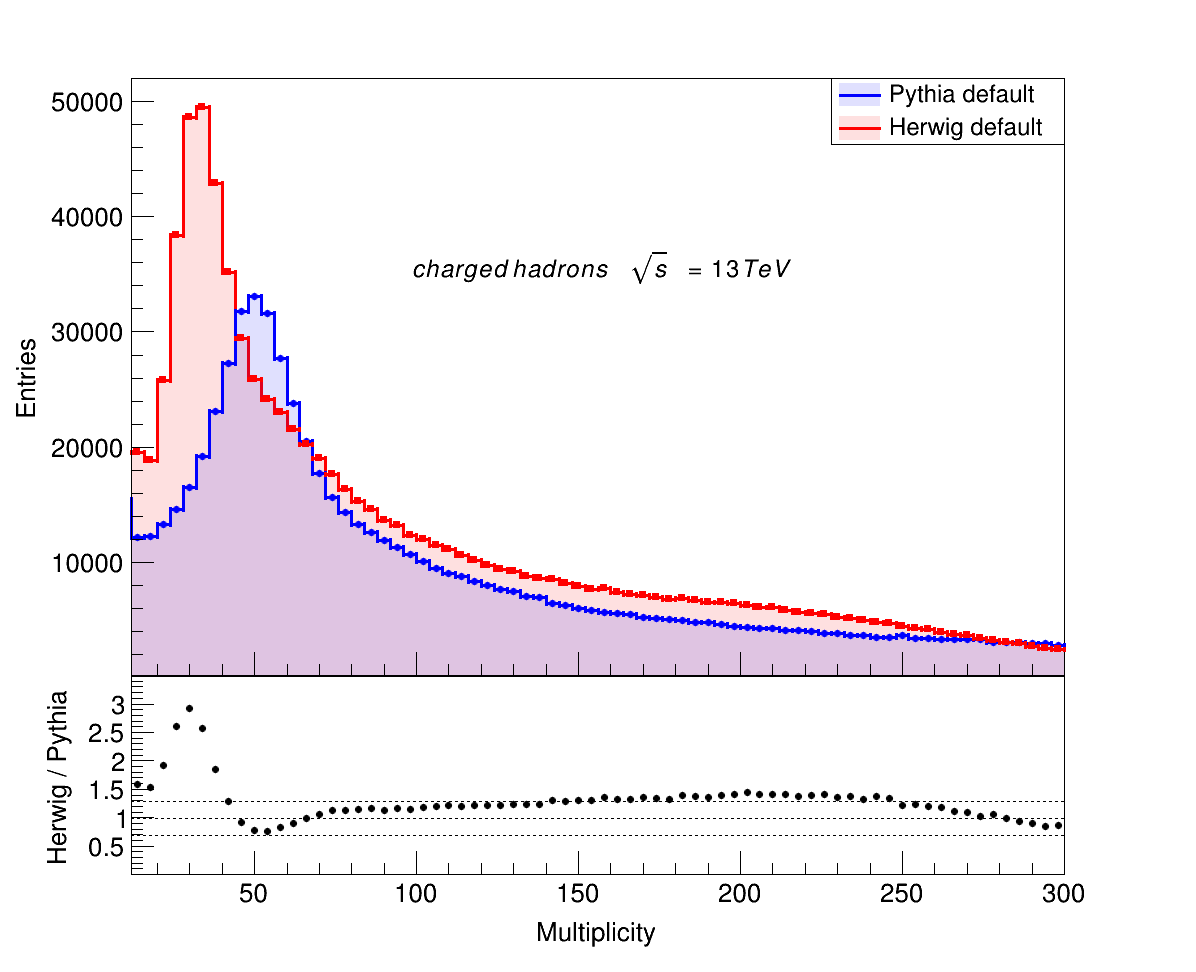}
    \caption{}
    \label{fig:mult_charged_13}
  \end{subfigure}
  \hspace{-1ex}
  \begin{subfigure}[b]{0.45\linewidth}
    \centering
    \includegraphics[width=\linewidth]{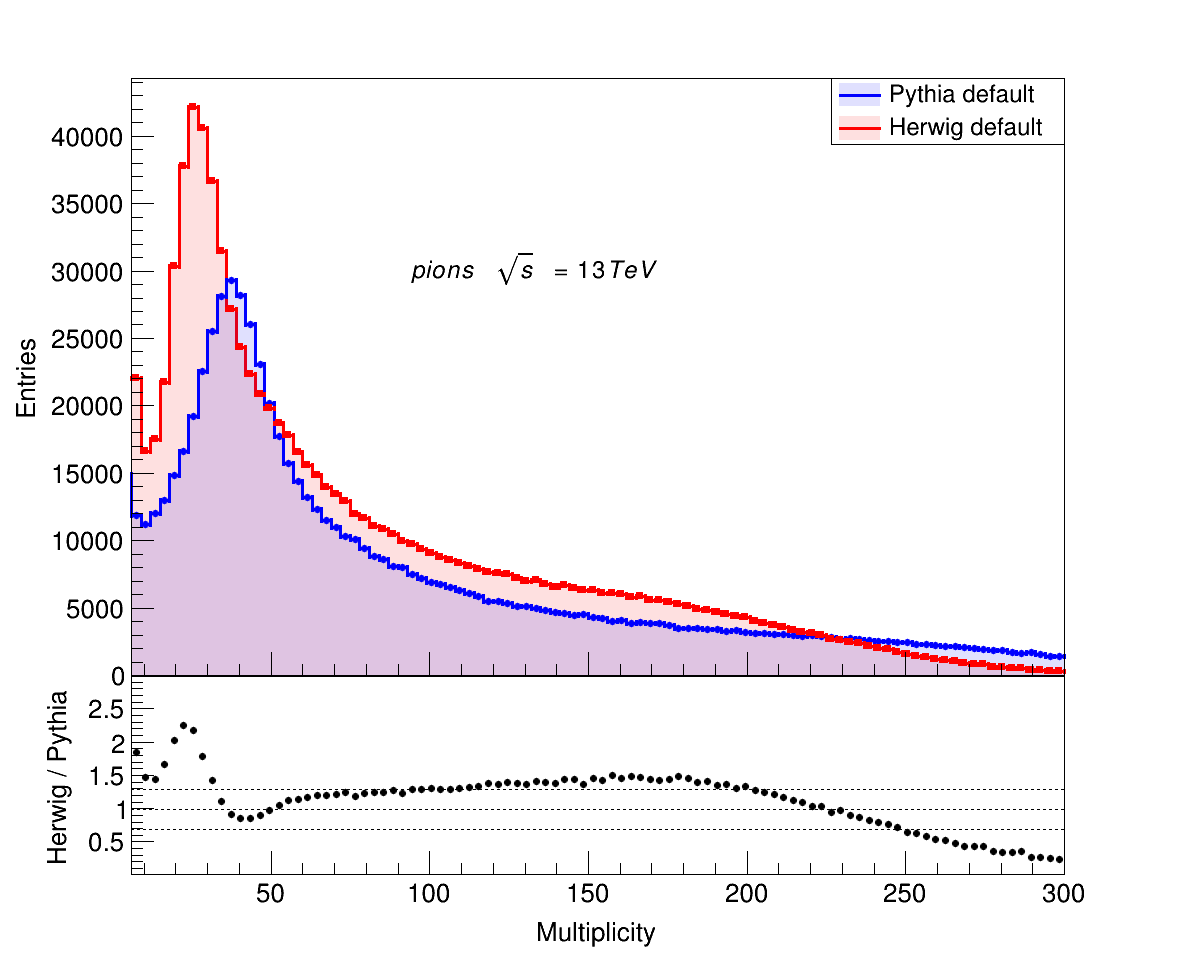}
    \caption{}
    \label{fig:mult_pions_13}
  \end{subfigure}

  \vspace{2ex}

  \begin{subfigure}[b]{0.45\linewidth}
    \centering
    \includegraphics[width=\linewidth]{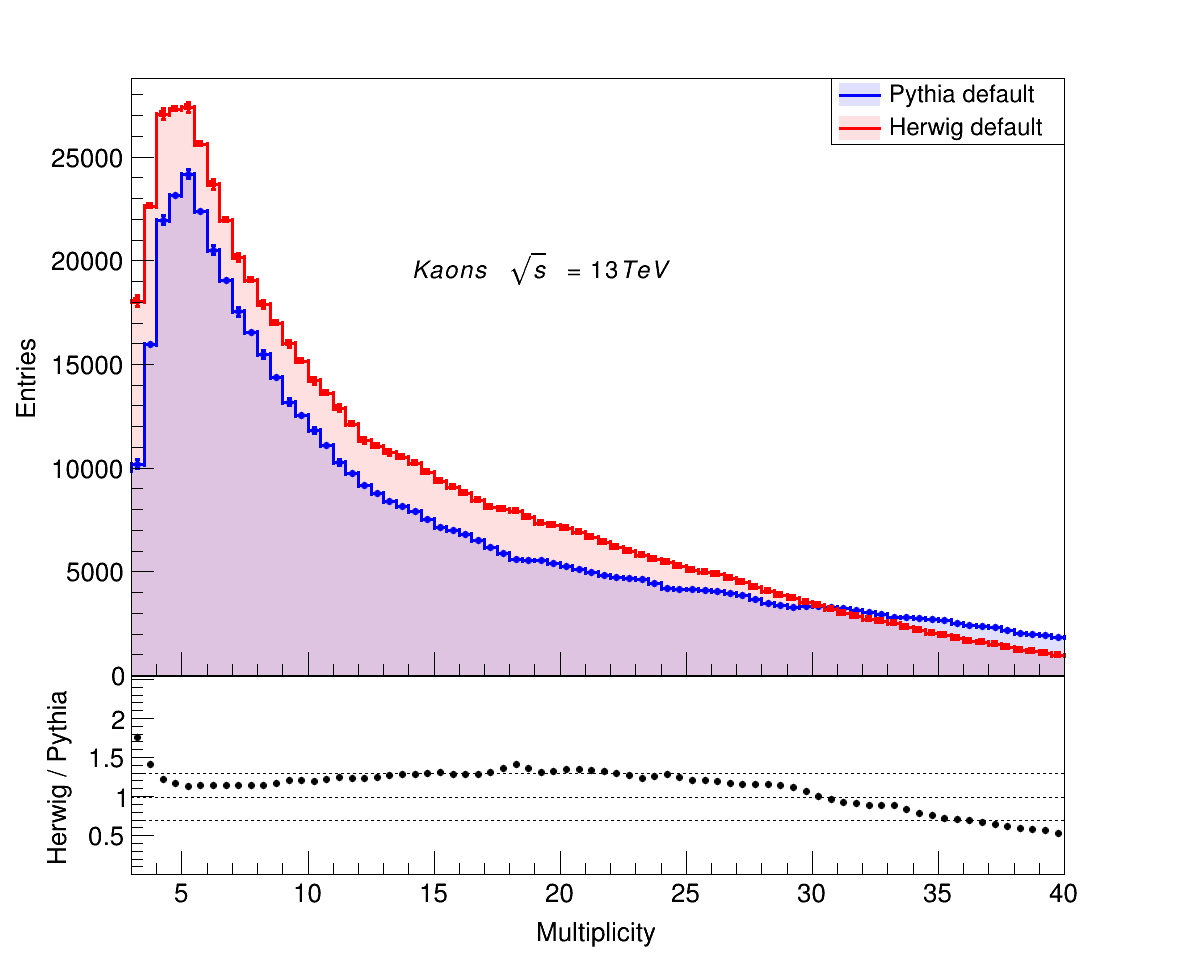}
    \caption{}
    \label{fig:mult_kaons_13}
  \end{subfigure}
  \hspace{-1ex}
  \begin{subfigure}[b]{0.45\linewidth}
    \centering
    \includegraphics[width=\linewidth]{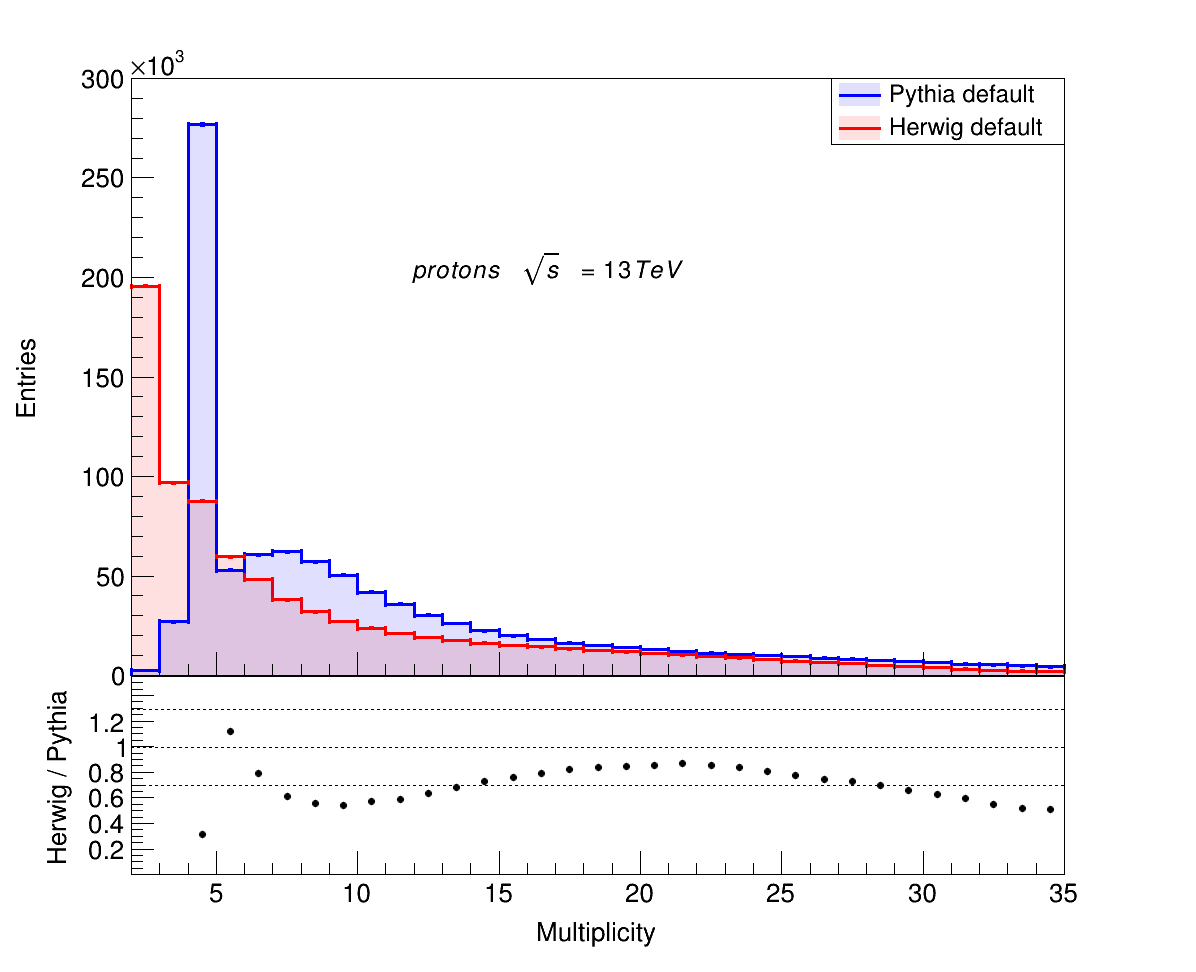}
    \caption{}
    \label{fig:mult_protons_13}
  \end{subfigure}

  \caption{Comparison of \textsc{Pythia} and \textsc{Herwig} multiplicity distributions at $\sqrt{s} = 13~\mathrm{TeV}$ for (a) charged hadrons, (b) pions, (c) kaons, and (d) protons. The error bars represent statistical uncertainties.}
  \label{pyth_herwig_comparison_mult13}
\end{figure*}


Finally, Fig.~\ref{pyth_herwig_comparison_mult7} shows the comparison between the two generators in the production of charged hadrons (a), pions (b), kaons (c), and protons (d) in an event at 7 TeV with statistical uncertainties. There are more charged particles produced in \textsc{Pythia}, as compared to \textsc{Herwig}. At low multiplicity, \textsc{Herwig} shows a higher particle production rate compared to \textsc{Pythia}. This is because \textsc{Herwig} tends to produce more particles in the initial stages of the parton shower process, which may be a result of differences in the underlying event and parton shower model. \textsc{Herwig} typically uses a cluster model for hadronization, where partons are clustered into hadrons in a way that often results in slightly more particles being produced initially. This can lead to higher multiplicity at low values. \textsc{Pythia}, on the other hand, often produces higher multiplicities at high energy due to its string model for hadronization. In this model, partons (quarks and gluons) create strings that stretch between them and fragment into hadrons, typically resulting in a higher number of final-state particles. MPI in \textsc{Pythia} can also significantly contribute to higher multiplicities, as it simulates several hard scatterings in a single event, allowing a broader spectrum of particle production at higher energies. A similar pattern could also be observed at 13 TeV as shown in Fig.~\ref{pyth_herwig_comparison_mult13}.

\begin{figure*}[!htbp]
  \captionsetup[subfigure]{font=scriptsize} 
  \centering

  \begin{subfigure}[b]{0.45\linewidth}
    \centering
    \includegraphics[width=\linewidth]{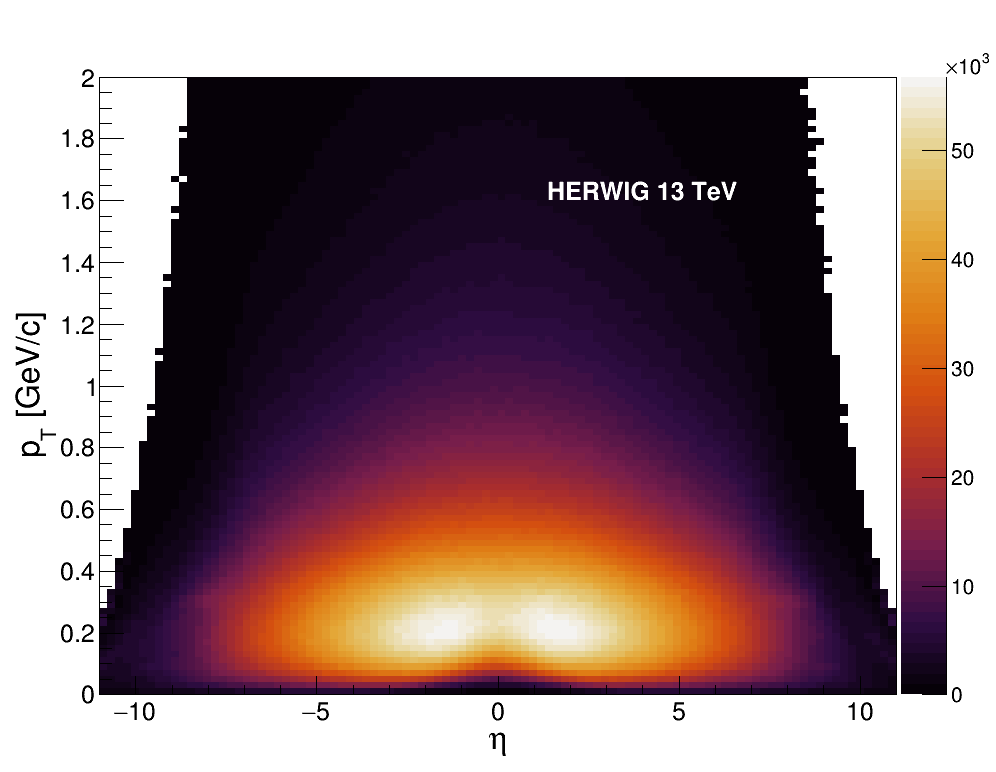}
    \caption{}
  \end{subfigure}
  \hspace{-1ex}
  \begin{subfigure}[b]{0.45\linewidth}
    \centering
    \includegraphics[width=\linewidth]{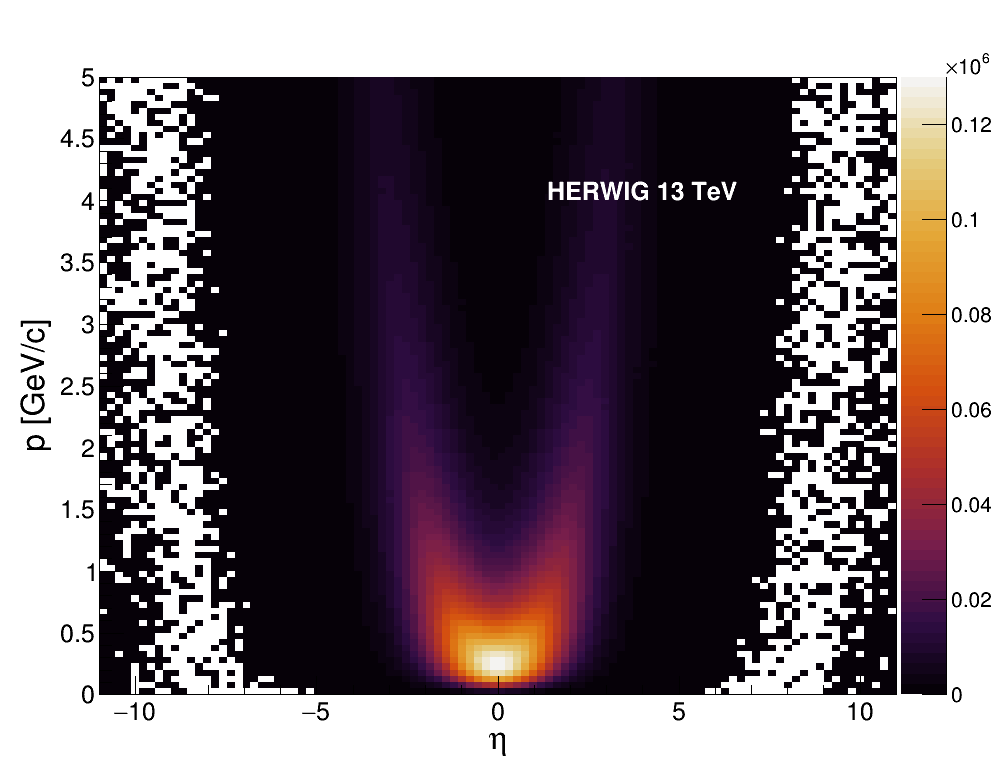}
    \caption{}
    \label{p_eta_herwig}
  \end{subfigure}

  \vspace{2ex}

  \begin{subfigure}[b]{0.45\linewidth}
    \centering
    \includegraphics[width=\linewidth]{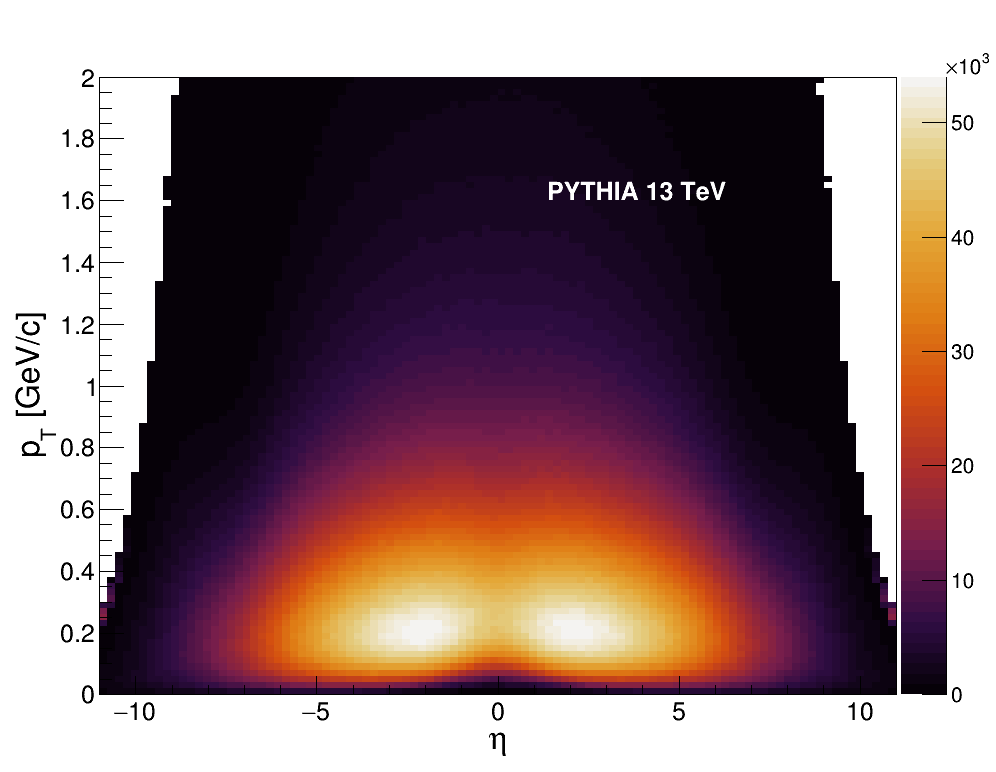}
    \caption{}
  \end{subfigure}
  \hspace{-1ex}
  \begin{subfigure}[b]{0.45\linewidth}
    \centering
    \includegraphics[width=\linewidth]{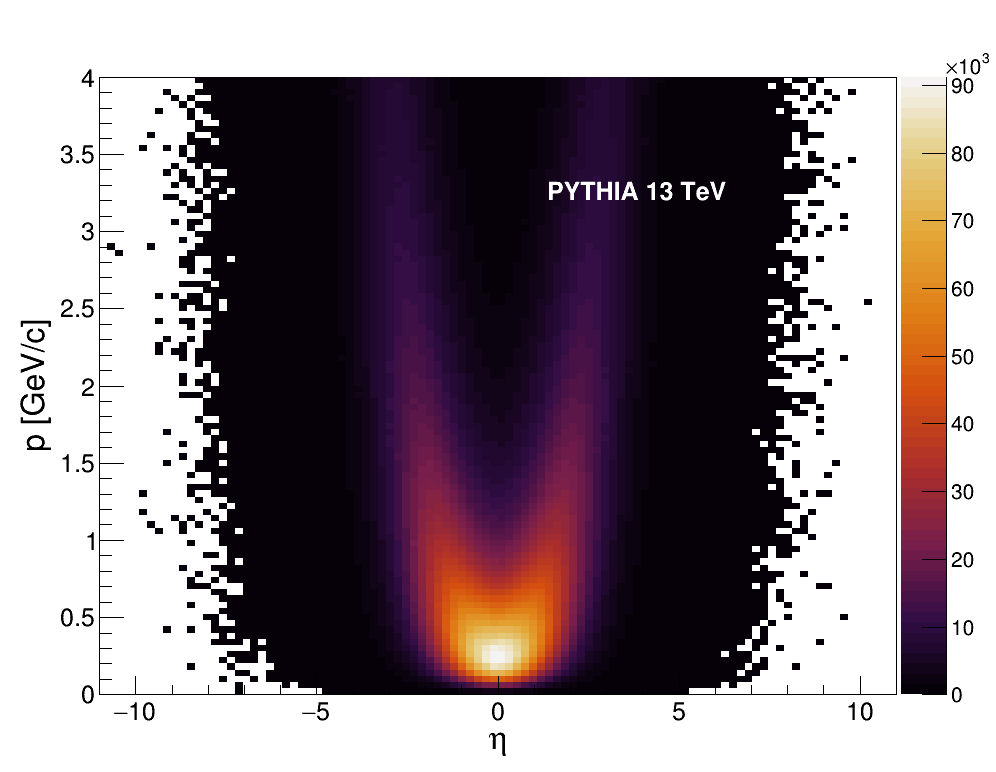}
    \caption{}
    \label{p_eta_pythia}
  \end{subfigure}

  \caption{Comparison of charged hadron distributions for \textsc{Herwig} and \textsc{Pythia} at $\sqrt{s} = 13~\mathrm{TeV}$. (a), (b) shows the $p_T$ and momentum versus $\eta$ in \textsc{Herwig} and (c), (d) shows $p_T$ and momentum versus $\eta$ in \textsc{Pythia}.}
  \label{2d_pt_p_eta}
\end{figure*}

Angular distribution of charged particles with respect to momentum and transverse momentum for \textsc{Pythia} and \textsc{Herwig} at 13 TeV is shown in Fig.~\ref{2d_pt_p_eta}. The resulting plots show a high concentration of low-momentum particles produced predominantly in the central regions, which could also be confirmed from \ref{pyth_herwig_comparison_eta13}. This behavior is consistent with expectations from soft QCD processes, where the majority of particles are produced with low transverse momenta and at mid-rapidity. Compared to \textsc{Herwig}, slight differences are observed in the shape and population density of \textsc{Pythia}. These differences arise from the distinct hadronization and parton shower models used in \textsc{Pythia}, which can lead to variations in particle multiplicity and kinematic distributions.

\begin{figure*}[!htbp]
  \captionsetup[subfigure]{font=scriptsize} 
  \centering

  \begin{subfigure}[b]{0.45\linewidth}
    \centering
    \includegraphics[width=\linewidth]{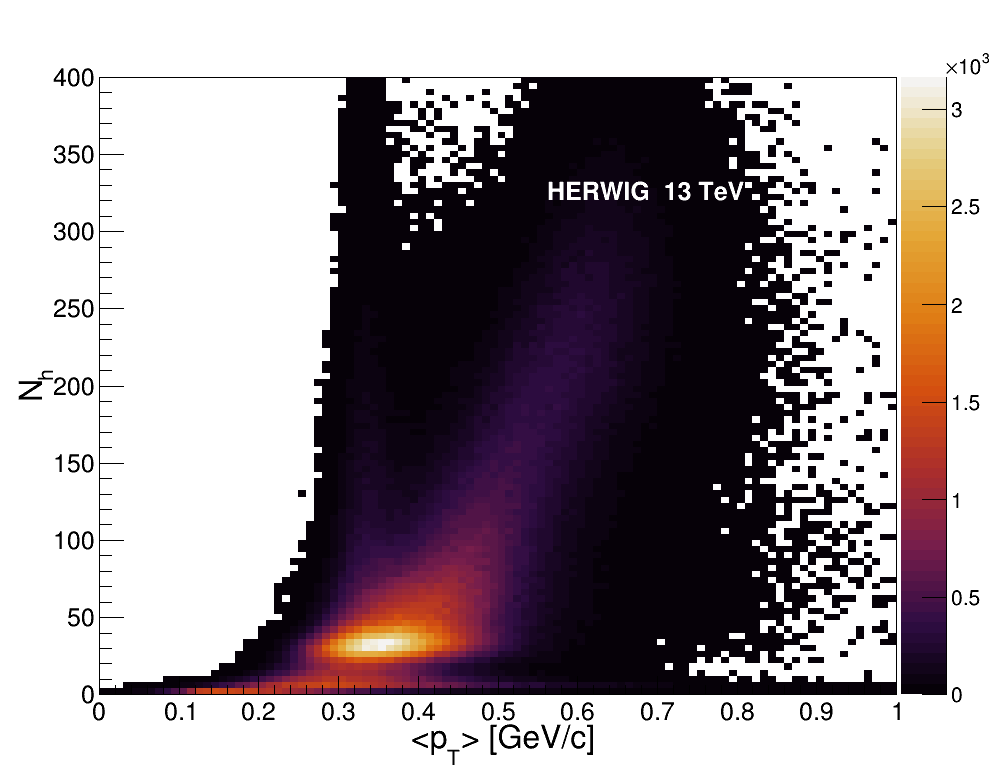}
    \caption{}
  \end{subfigure}
  \hspace{-1ex}
  \begin{subfigure}[b]{0.45\linewidth}
    \centering
    \includegraphics[width=\linewidth]{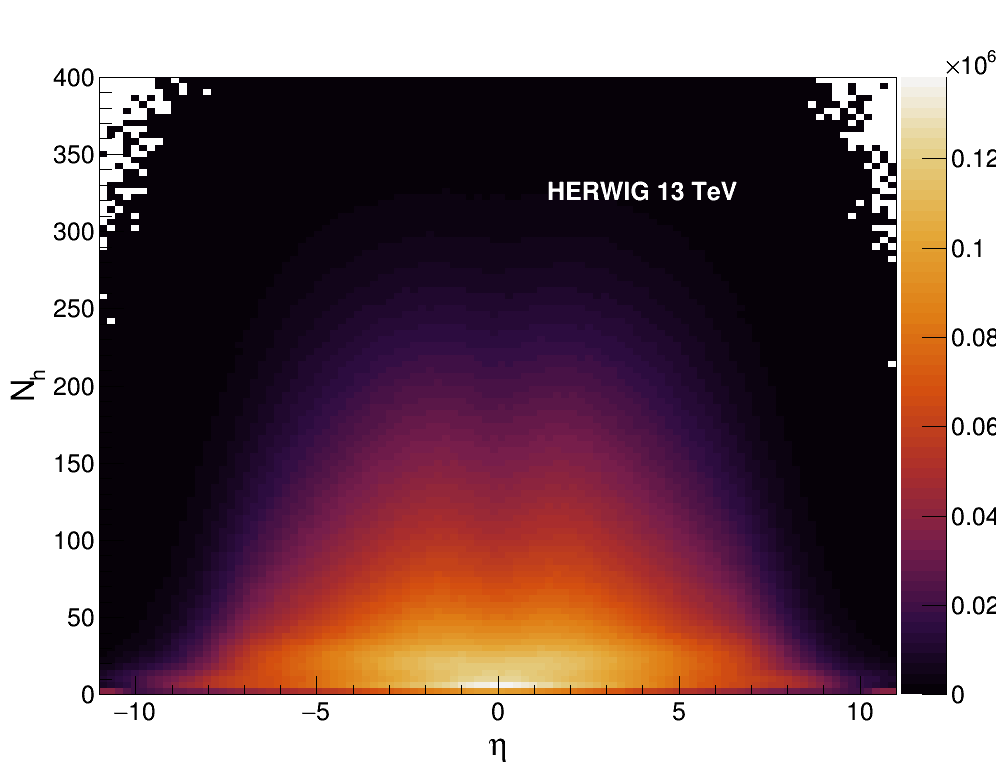}
    \caption{}
    \label{mult2Deta_charged13H}
  \end{subfigure}

  \vspace{2ex}

  \begin{subfigure}[b]{0.45\linewidth}
    \centering
    \includegraphics[width=\linewidth]{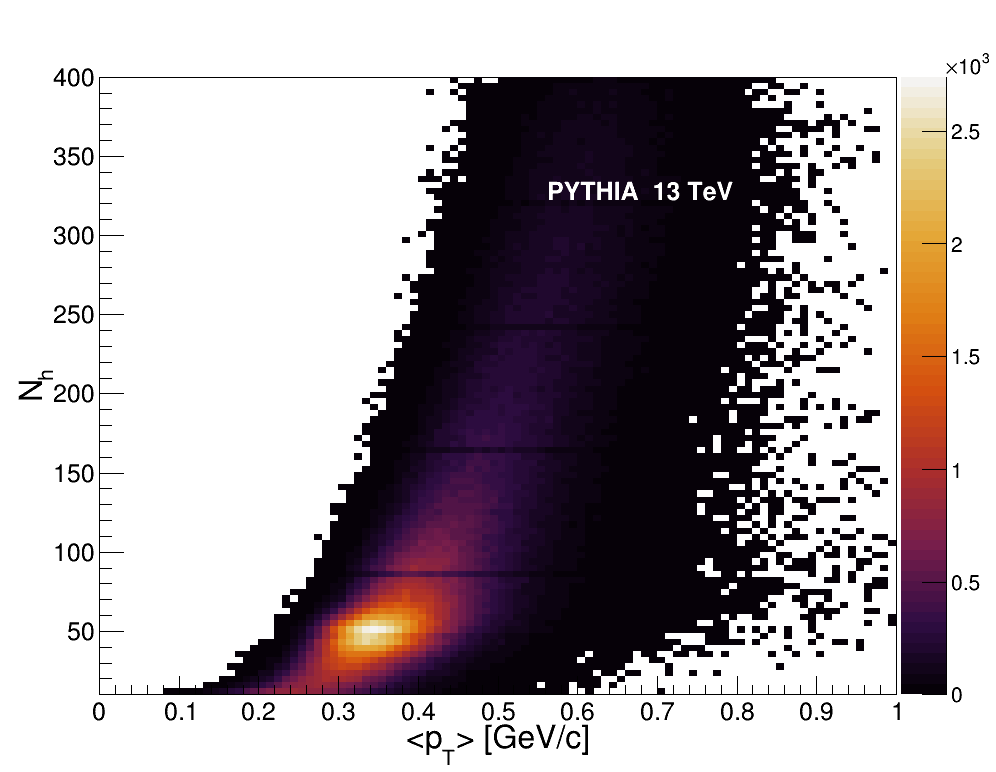}
    \caption{}
  \end{subfigure}
  \hspace{-1ex}
  \begin{subfigure}[b]{0.45\linewidth}
    \centering
    \includegraphics[width=\linewidth]{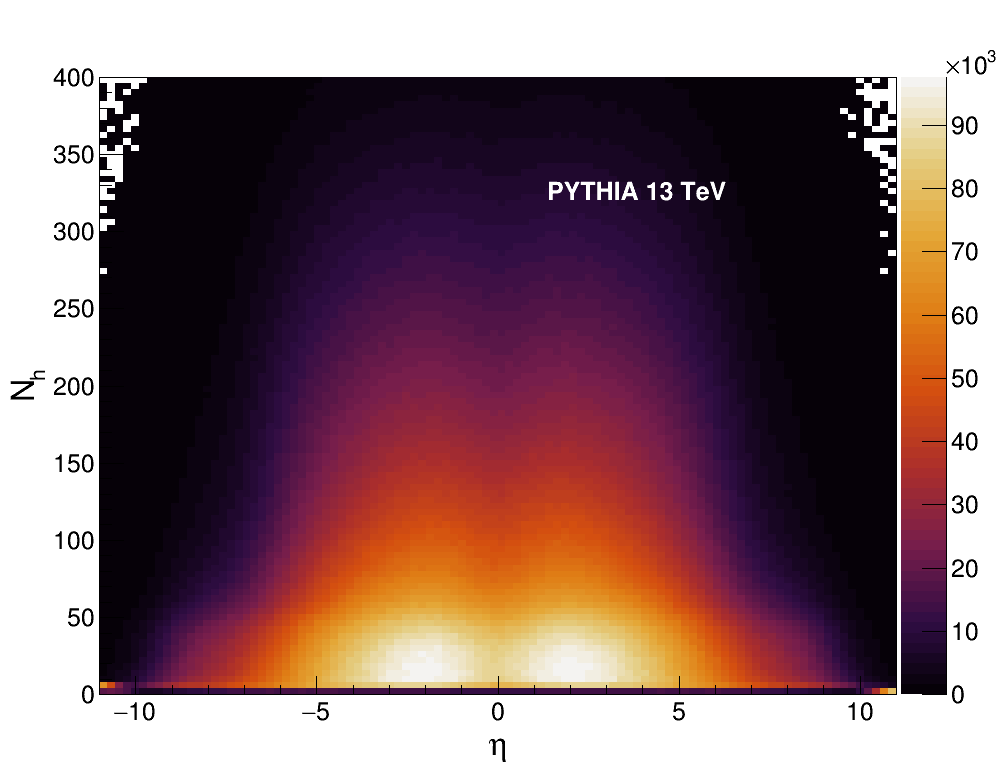}
    \caption{}
    \label{mult2Deta_charged13P}
  \end{subfigure}

  \caption{Comparison of charged hadron distributions for \textsc{Herwig} and \textsc{Pythia} at $\sqrt{s} = 13~\mathrm{TeV}$. (a) and (c) show the relationship between multiplicity and mean $p_T$ of charged hadrons, and (b) and (d) show the relationship between multiplicity and $\eta$ of charged hadrons.}
  \label{2d}
\end{figure*}

Figure~\ref{2d} shows the charged hadrons multiplicity with respect to momentum and mean transverse momentum for \textsc{Herwig} and \textsc{Pythia} as shown in Fig.~\ref{2d} (a) and (c). For the \textsc{Herwig} distribution, the mean $p_T$ shows a sharp peak at low $p_T$ values (~0.4 GeV/c) for events with moderate multiplicities followed by a rapid drop-off, indicating that most events are soft and low in hadronic activity consistent with \textsc{Herwig}'s modelling of soft interactions, whereas, for \textsc{Pythia}, there are slightly higher values of mean $p_{T}$ observed for events with moderate multiplicities. 

\begin{figure}
    \centering
    \includegraphics[width=0.65\linewidth]{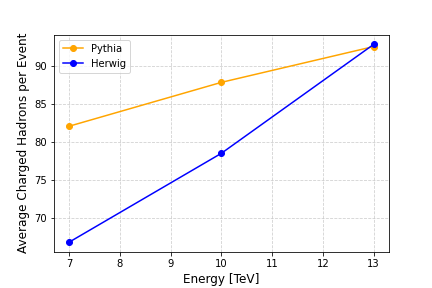}
    \caption{Energy dependence of the average charged-particle production in $pp$ collisions as simulated with \textsc{Pythia} and \textsc{Herwig} at $\sqrt{s} = 7$, $10$, and $13~\mathrm{TeV}$.}
    \label{energy_dependence}
\end{figure}

On the other hand, \textsc{Pythia}'s and \textsc{Herwig}'s multiplicity vs. $\eta$ distribution displays a broad, symmetric structure centered at $\eta$ = 0, with the highest event densities concentrated at low multiplicities and central rapidities. The comparison between \textsc{Herwig} and \textsc{Pythia} highlights key differences in their underlying physics models, especially in how they generate high-multiplicity or forward events. Such variations are particularly relevant when considering detector acceptance and efficiency in regions like those covered by LHCb. Furthermore, the charged-particle multiplicity predicted by both \textsc{Pythia} and \textsc{Herwig} exhibits a clear dependence on the center-of-mass energy, as illustrated in Fig.\ref{energy_dependence}, with notable variations observed at $\sqrt{s}$ = 7, 10, and 13~TeV.


\section{Impact of modification of parameters on particle production}

The importance of parameters becomes particularly evident when analyzing the plots. Even slight changes in parameter values can have a significant impact on key observables such as transverse momentum, pseudorapidity, momentum, and particle production rates. To understand how these parameters influence the results, we refer to Eq.~\ref{factorisation}. Among various parameters, the most critical ones are $p^{T0}_{ref}$, $\epsilon$ and $\alpha_{s}$. These parameters, as defined earlier, collectively govern the behavior of the system and shape the distributions of the physical quantities being studied. 

\begin{figure*}[h]
    \centering
    \begin{subfigure}[b]{0.45\linewidth}
        \centering
        \includegraphics[width=\linewidth]{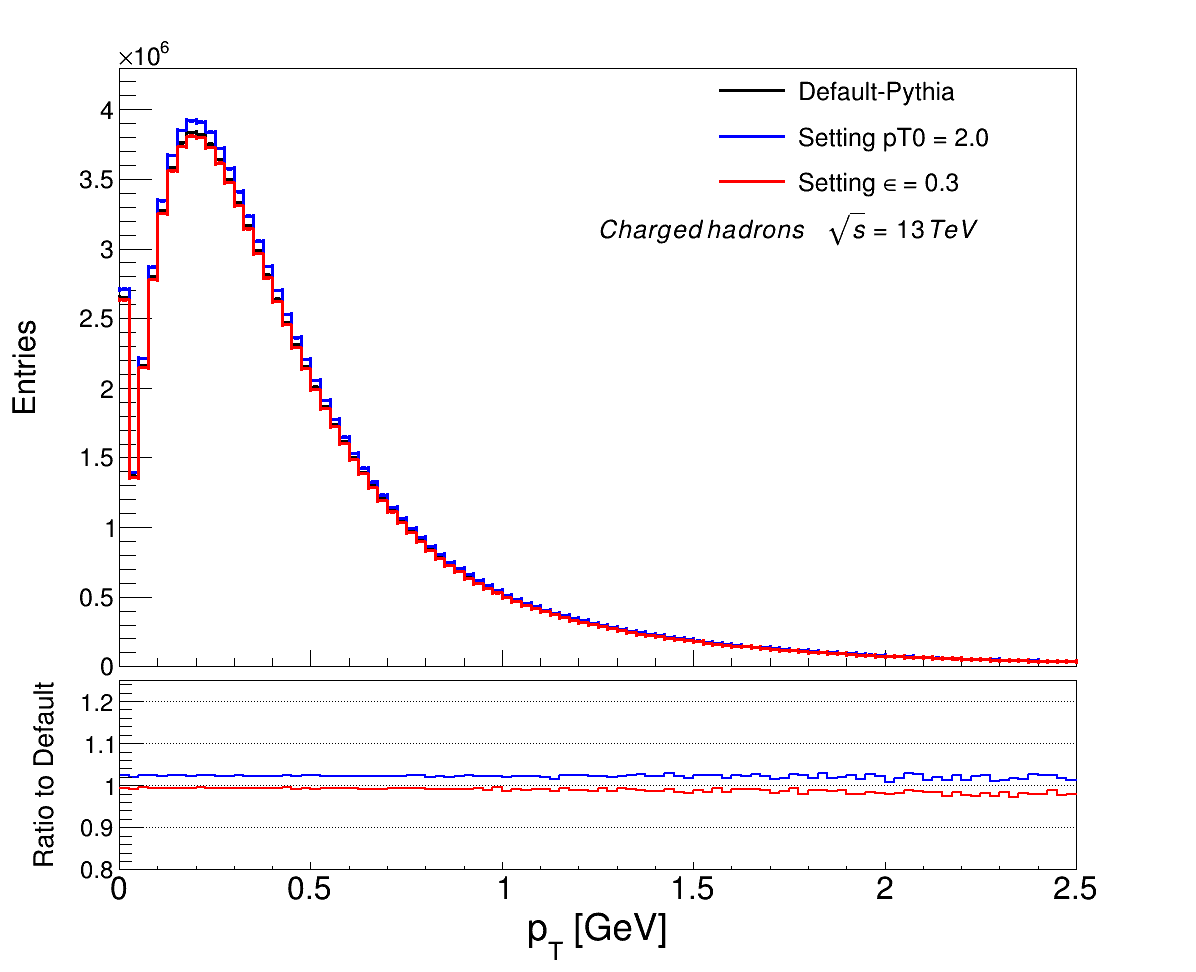}
        \caption{}
    \end{subfigure}
    \hspace{0.01\linewidth} 
    \begin{subfigure}[b]{0.45\linewidth}  
        \centering 
        \includegraphics[width=\linewidth]{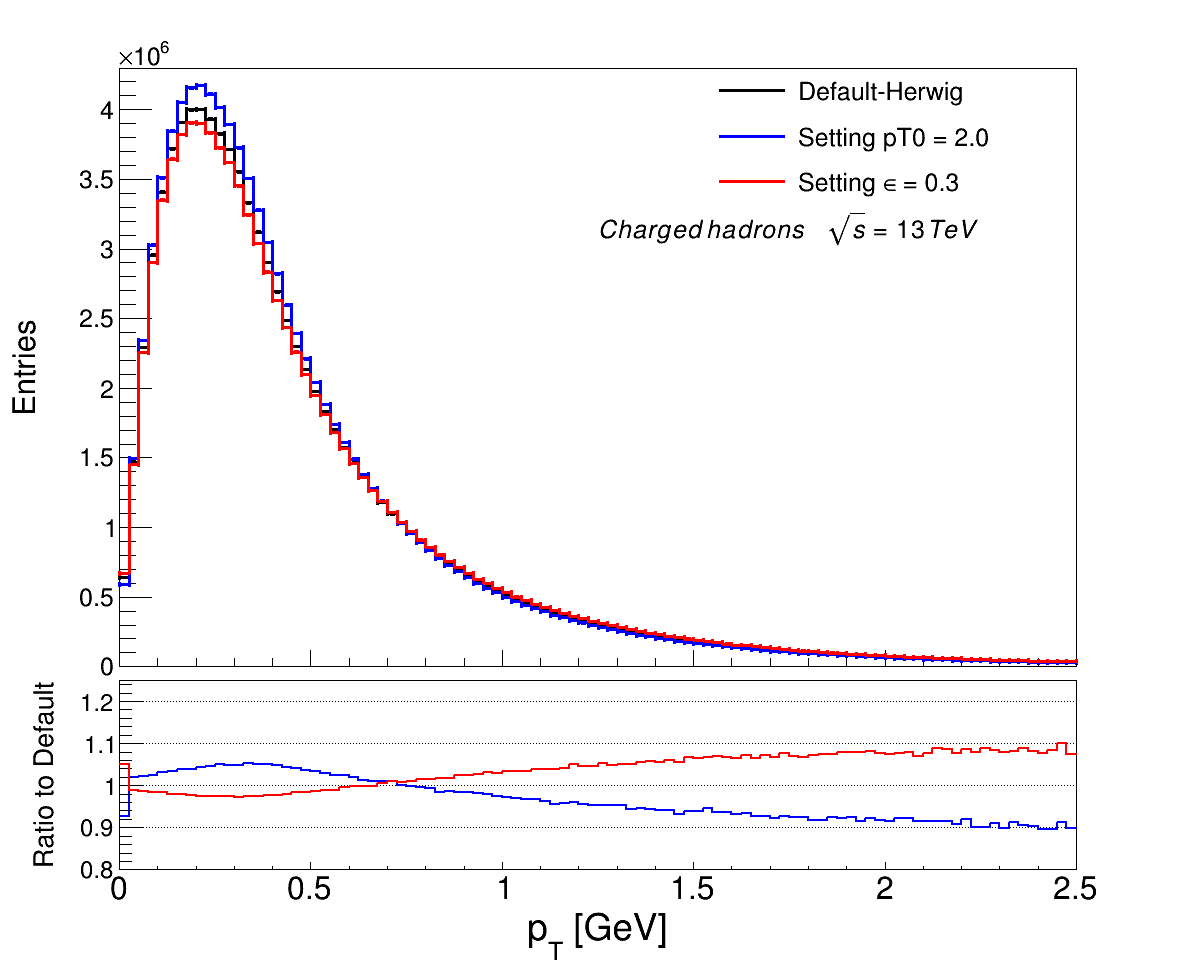}
        \caption{}
    \end{subfigure}
    \caption{Influence of modifying parameters of \textsc{Pythia} (a) and \textsc{Herwig} (b) in the transverse momentum of charged hadrons. The error bars represent statistical uncertainties.}
    \label{pythia_herwig_compare_all_pt}
\end{figure*}

\begin{figure*}[h]
    \centering
    \begin{subfigure}[b]{0.45\linewidth}
        \centering
        \includegraphics[width=\linewidth]{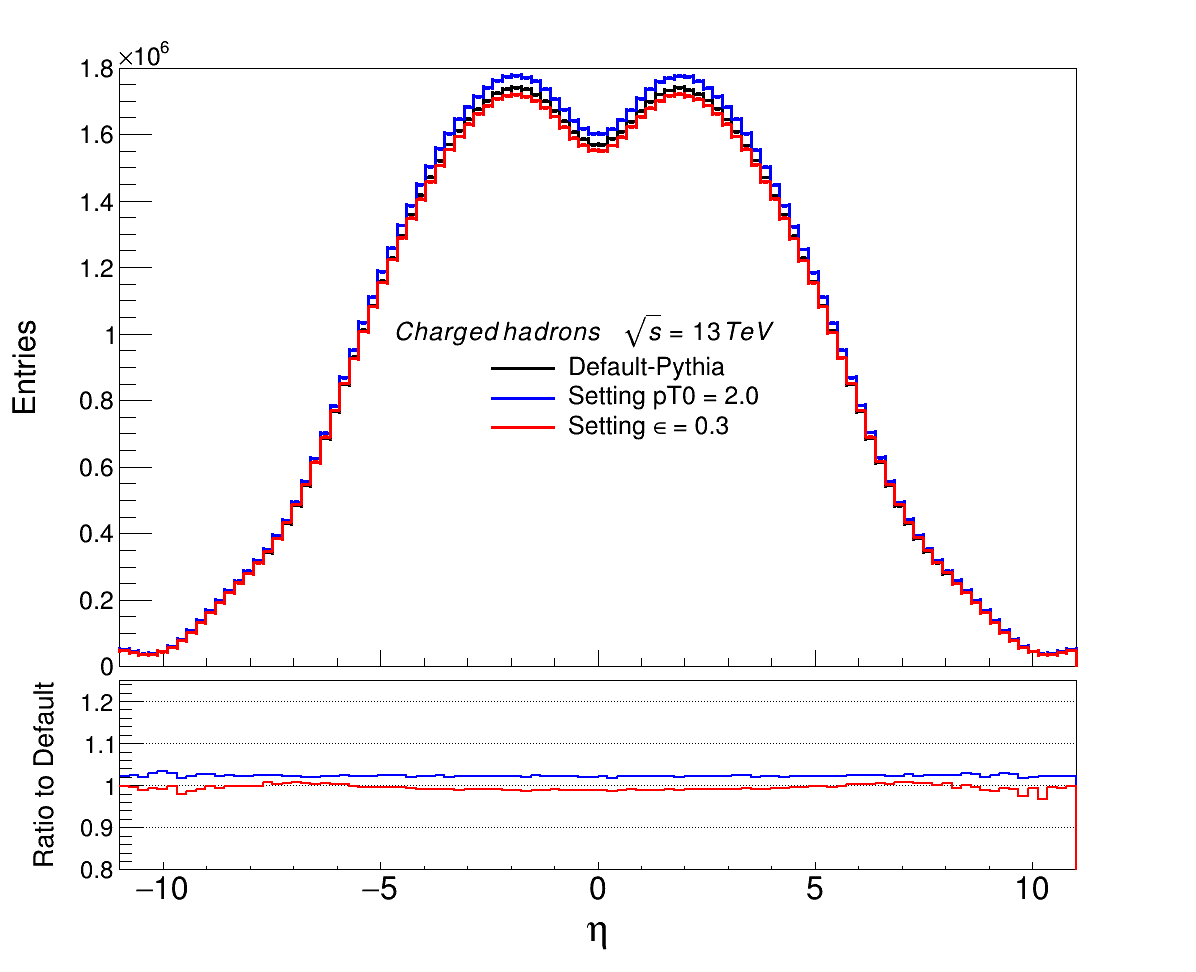}
        \caption{}
    \end{subfigure}
    \hspace{0.01\linewidth} 
    \begin{subfigure}[b]{0.45\linewidth}  
        \centering 
        \includegraphics[width=\linewidth]{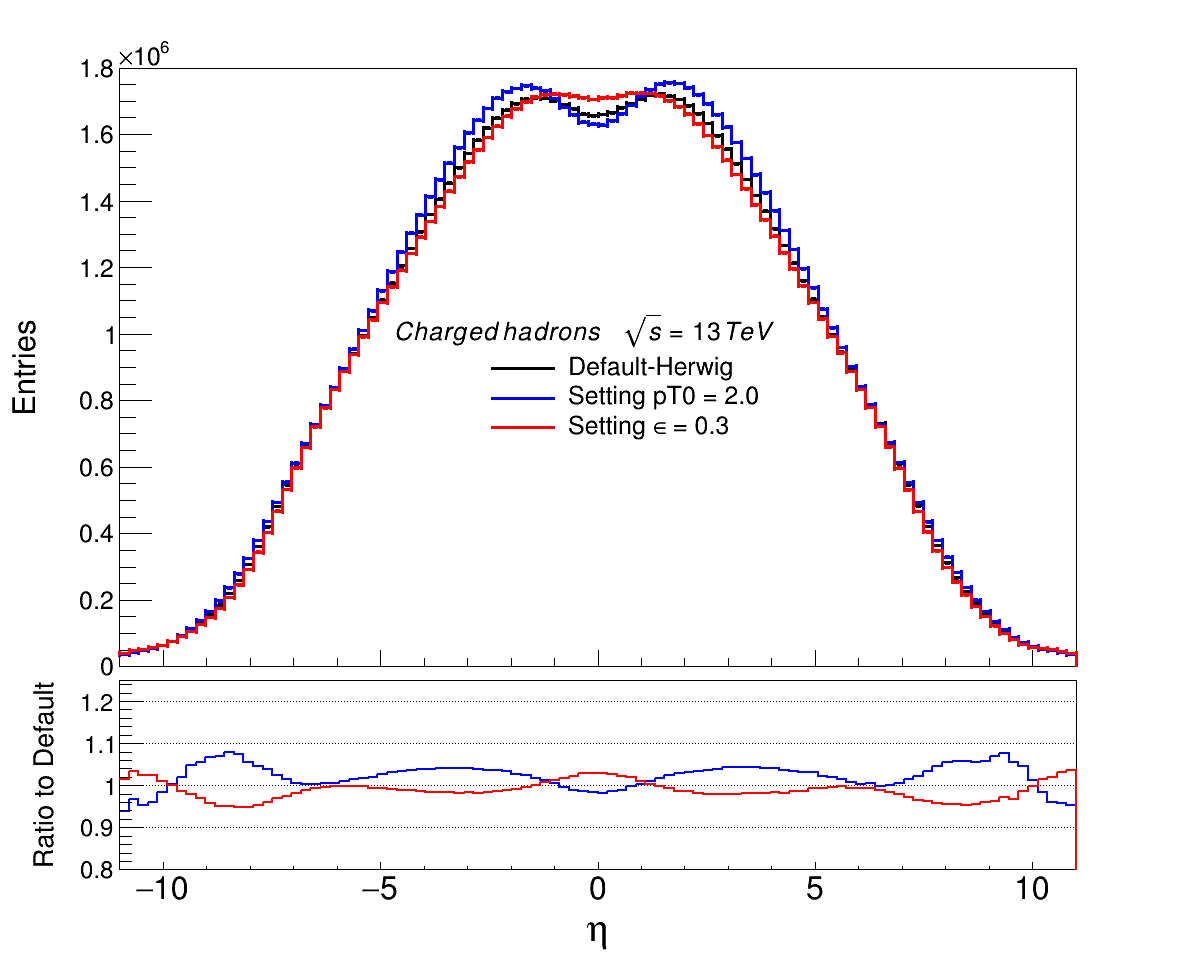}
        \caption{}
    \end{subfigure}
    \caption{Influence of modifying parameters of \textsc{Pythia} (a) and \textsc{Herwig} (b) as a function of $\eta$ of charged hadrons. The error bars represent statistical uncertainties.}
    \label{pythia_herwig_compare_all_eta}
\end{figure*}

\begin{figure*}[h]
    \centering
    \begin{subfigure}[b]{0.45\linewidth}
        \centering
        \includegraphics[width=\linewidth]{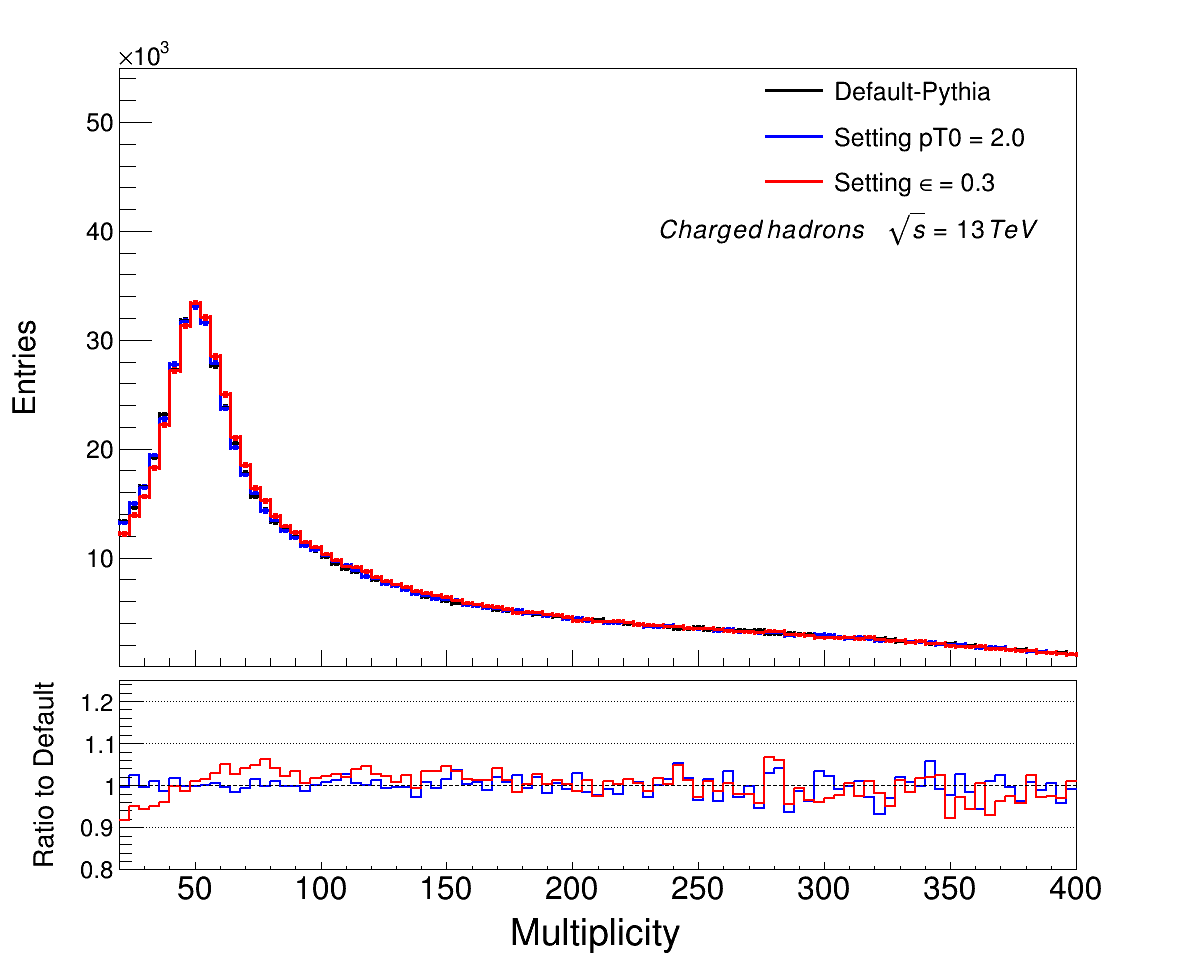}
        \caption{}
    \end{subfigure}
    \hspace{0.01\linewidth} 
    \begin{subfigure}[b]{0.45\linewidth}  
        \centering 
        \includegraphics[width=\linewidth]{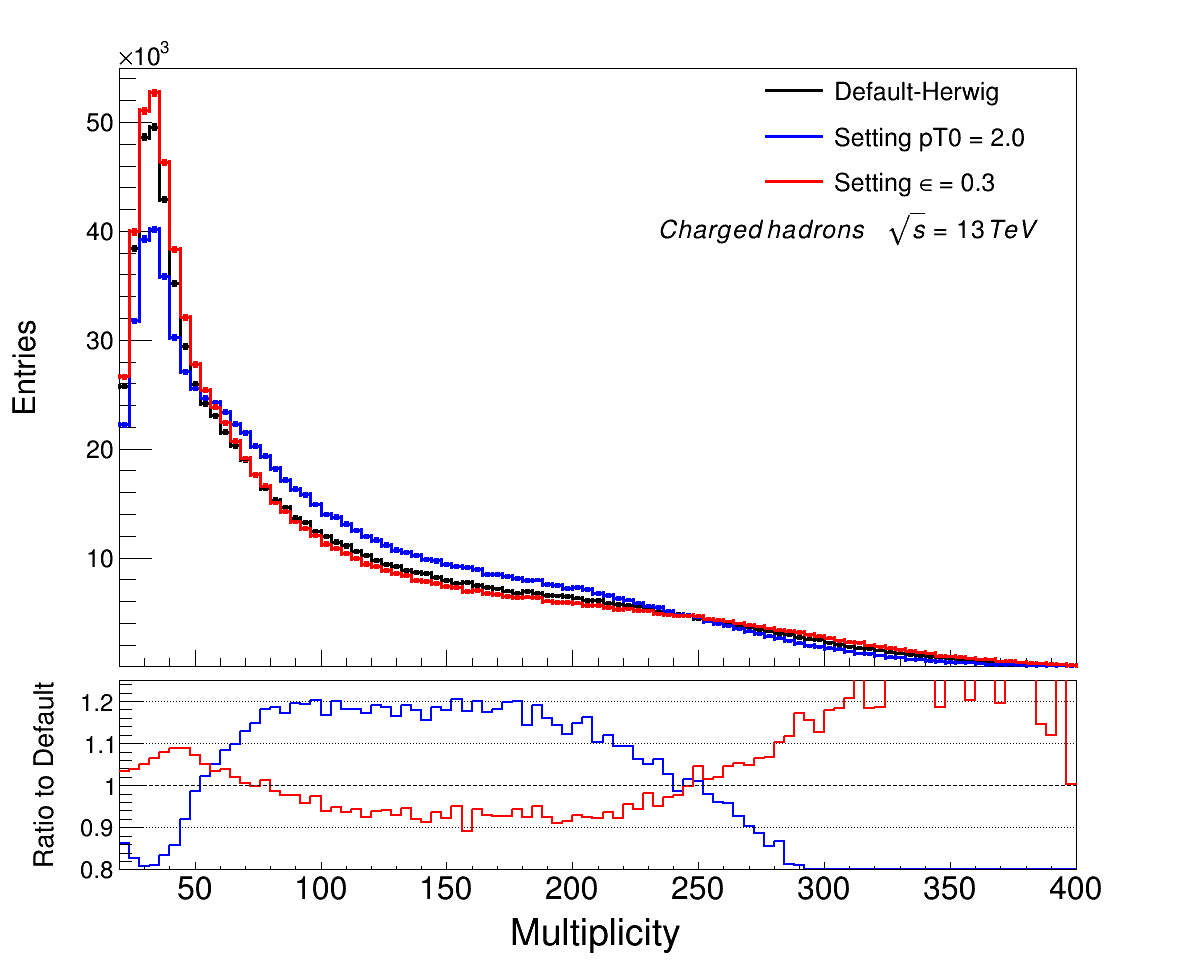}
        \caption{}
    \end{subfigure}
    \caption{Influence of modifying parameters of \textsc{Pythia} (a) and \textsc{Herwig} (b) as a function of multiplicity of charged hadrons. The error bars represent statistical uncertainties.}
    \label{pythia_herwig_compare_all_mult}
\end{figure*}

To observe how parameters influence the distributions, Figs.~\ref{pythia_herwig_compare_all_pt}, \ref{pythia_herwig_compare_all_eta}, and \ref{pythia_herwig_compare_all_mult} show a comparison of \textsc{Pythia} (a) and \textsc{Herwig} (b) for transverse momentum, $\eta$ and multiplicity with respect to the default and two modified settings in the values of parameters. The plots show three distributions, the solid black line refers to the default distribution, whereas the solid blue shows when the value of the cut-off parameter $p_{T0}^{ref}$ is reduced from the default to 2.0 and the red line shows when the scaling parameter $\epsilon$ is changed to 0.3 for \textsc{Pythia} and \textsc{Herwig}. As mentioned earlier, $p_{T0}^{ref}$ acts like a barrier between hard and soft interactions; therefore, once the value is reduced, there are more soft interactions produced, causing an increase in multiplicity. Since the value of the default parameter for $p_{T0}^{ref}$ is 2.28 for \textsc{Pythia} and 3.1 for \textsc{Herwig}, reducing this value in \textsc{Pythia} won't show a significant impact on the number of particles, whereas the difference is visible in the $p_{T}$ distribution of \textsc{Herwig}. 

The other important parameter that is modified is $\epsilon$, i.e., the exponent of Eq.~\ref{pt0} that controls the energy scaling behavior of the cut-off parameter and is crucial for ensuring that soft or low-energy emissions do not cause divergences in the perturbative QCD calculations. The comparison of \textsc{Pythia} and \textsc{Herwig} for the charged hadrons with respect to transverse momentum is shown in Fig.~\ref{pythia_herwig_compare_all_pt}.
However, the distributions show that this parameter does not have a drastic impact. For \textsc{Pythia}, the distribution is quite similar to the default; however, a slight change is visible at the lower $p_{T}$ values in \textsc{Herwig}.


Figure~\ref{pythia_herwig_compare_all_eta}, shows the $\eta$ distribution for \textsc{Pythia} in (a) and \textsc{Herwig} in (b). The distribution is roughly symmetric around $\eta$ = 0, consistent with the symmetry of proton-proton collisions, showing that the entries are concentrated near $\eta$ = 0 corresponding to particles emitted perpendicular to the beam axis, with fewer particles at large $\eta$ where particles travel closer to the beamline. The modifications have more noticeable effects compared to \textsc{Herwig}, particularly between $\eta = \pm 5$.

The multiplicity plots for both generators are presented in Fig.~\ref{pythia_herwig_compare_all_mult}. For \textsc{Pythia}, the multiplicity demonstrates minimal variation with parameter changes, as the differences in the default values and modified values are relatively small. In contrast, \textsc{Herwig} exhibits a more noticeable variation. Detailed numerical values are provided in Table \ref{average_modified_hadrons13}, offering a precise representation of the particle count per event as seen in the plots.

\begin{table*}[ht]
\small
\centering
\begin{tabular}{| *{9}{c|} }
    \hline
    & \multicolumn{2}{c|}{$N_{\pi}$}
    & \multicolumn{2}{c|}{$N_{K}$}
    & \multicolumn{2}{c|}{$N_{p}$}
    & \multicolumn{1}{c|}{$N_{ch}$} \\
    \hline
\textsc{Pythia} (p$_{T0}^{ref}$ = 2.0 GeV) & 71.66 $\pm$ 0.083 & 77.5\%
& 9.68 $\pm$ 0.012 & 10.47\%
& 11.077 $\pm$ 0.009 & 11.98\%
& 92.42 $\pm$ 0.103 \\
\hline
\textsc{Herwig} (p$_{T0}^{ref}$ = 2.0 GeV) & 77.31 $\pm$ 0.064 & 81.55\%
& 9.84 $\pm$ 0.009 & 10.38\%
& 7.61 $\pm$ 0.007 & 8.02\%
& 94.78 $\pm$ 0.079 \\
\hline
\textsc{Pythia} ($\epsilon$ = 0.3) & 71.62 $\pm$ 0.082 & 77.5\%
& 9.68 $\pm$ 0.012 & 10.47\%
& 11.075 $\pm$ 0.009 & 11.98\%
& 92.38 $\pm$ 0.102 \\
\hline
\textsc{Herwig} ($\epsilon$ = 0.3) & 74.36 $\pm$ 0.068 & 80.64\%
& 9.78 $\pm$ 0.010 & 10.60\%
& 8.06 $\pm$ 0.009 & 8.74\%
& 92.21 $\pm$ 0.085 \\
\hline
\end{tabular}
\caption{Average number of charged hadrons in an event for the modified settings of \textsc{Pythia} and \textsc{Herwig} at $\sqrt{s} = 13$ TeV with statistical uncertainties. The percentage indicates its fraction relative to the total average number of charged hadrons $N_{ch}$.}
\label{average_modified_hadrons13}
\end{table*}

The parameters are extremely sensitive; small variations can substantially affect the results. Therefore, parameter selection must be approached with caution. Additionally, to determine which generator best describes the experimental data, a comparison within the LHCb acceptance regions and corresponding LHCb data is necessary and is discussed in the next section.

Systematic theoretical uncertainties are assessed through variations of key generator
parameters, following the standard procedure adopted in LHCb Monte Carlo studies
and in the generator–tuning community
(e.g.\ Refs.~\cite{pythia_8.1,Skands_tuningMCEG}).
The parameters listed in Tables~1 and~2
($p_{T0}^{\mathrm{ref}}$, $\epsilon$, $\alpha_s$, etc.)
are tuned quantities rather than fixed theoretical constants. Although explicit uncertainties are not always quoted in the generator manuals, it is standard practice to estimate their impact by varying them within physically reasonable ranges. In the present work, we vary
$p_{T0}^{\mathrm{ref}}$ by $\pm(10$--$20)\%$, $\epsilon$ by $\pm0.02$, and $\alpha_s$ by $\pm0.001$--$0.002$. For \textsc{Herwig}, the present work represents an initial exploratory comparison.
A full systematic variation campaign has not yet been performed,
but the same parameter variation approach will be used in future
tune–validation studies to quantify theoretical uncertainties.


\section{Comparison of models in \textsc{Pythia} and \textsc{Herwig} with LHCb settings}
\label{comp_lhcb}

Event generation at LHCb is predominantly carried out with \textsc{Pythia}~\cite{simproject}. This section makes a comparison between \textsc{Pythia} and \textsc{Herwig}, taking into account the LHCb data, which is taken from analysis \cite{mult_comparison}, and for the convenience of the comparison, the \textsc{Rivet} analysis based on the data is used. \textsc{Rivet}~\cite{rivet} is a software tool used to easily compare theoretical particle physics generators with experimental data from particle collisions, such as those studied at LHCb. This helps to validate and tune theoretical models by providing a standardized way to evaluate event-level data without biasing the results to specific detector effects. 

To evaluate the models, the LHCb measurement of charged particle multiplicity and density~\cite{mult_comparison} in $pp$ collisions at 7 TeV with a low interaction rate provides an excellent test. In this case, visible events at the generator level are required to contain at least one charged particle within the pseudorapidity range $2.0 < \eta < 4.8$ and with transverse momentum $p_{T} > 0.2$ GeV, a momentum of $p > 2 $ GeV, and a lifetime of $\tau < 10$ ps. A reconstructed event must contain at least one track transversing all LHCb tracking stations as well as passing within 2 mm of the beam line and originating from the luminous region of the collision. 

The results from LHCb are compared to simulations from \textsc{Pythia} and \textsc{Herwig} using the tuned parameters set by LHCb for 7 TeV data as listed in Table~\ref{lhcb_tunedsettings} and mentioned in~\cite{lhcb_pythia_7tev_settings} and~\cite{lhcb_herwig_7tev_settings} for \textsc{Pythia} and \textsc{Herwig}, respectively. The available “LHCb” tune in this version is derived from the general LHC underlying-event tune, with adjustments based on comparisons to LHCb phase-space observables.

\begin{figure*}[!htbp]
    \centering
    \begin{subfigure}{0.45\linewidth}
        \includegraphics[width=\textwidth]{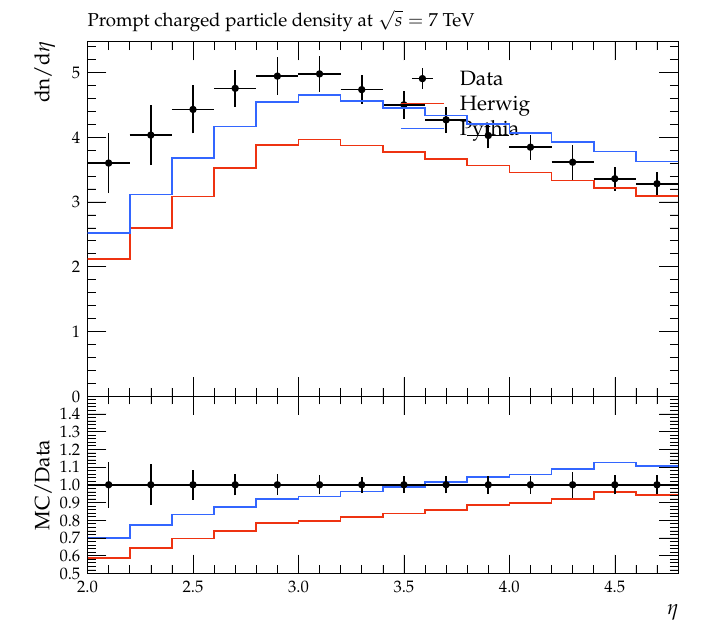}
        \caption{}
        \label{a}
    \end{subfigure}
    \hspace{0.05\linewidth}
    \begin{subfigure}{0.45\linewidth}
        \includegraphics[width=\textwidth]{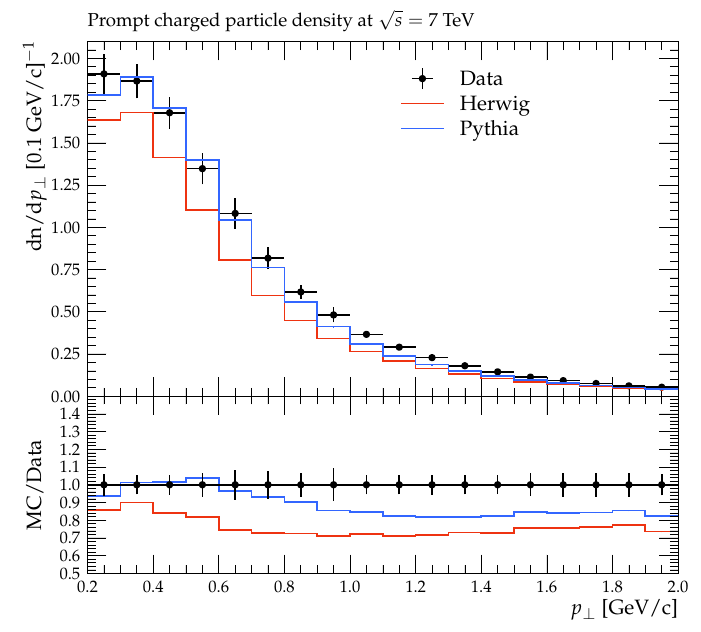}
        \caption{}
        \label{b}
    \end{subfigure}
    \vspace{1em} 
    \begin{subfigure}{0.45\linewidth}
        \centering
        \includegraphics[width=\textwidth]{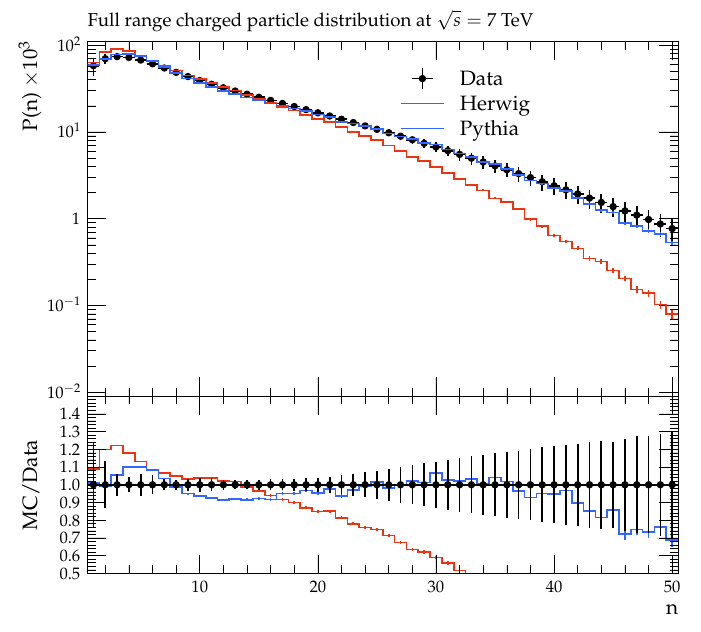}
        \caption{}
        \label{c}
    \end{subfigure}
    \caption{Charged particle density as a function of pseudorapidity $\eta$ in (a), $p_{T}$ in (b) and of charged particle multiplicity in (c) at $\sqrt{s} = 7$ TeV for \textsc{Pythia} and \textsc{Herwig} in solid lines compared with the LHCb data represented by the points.}
    \label{all_figs}
\end{figure*}

The plots in Fig.~\ref{all_figs} show the differential charged particle density as a function of $\eta$ (a), and $p_{T}$ (b), and the full kinematic range (entire phase space) charged particle distribution (c). 

\begin{table}[ht]
\small
\setlength{\tabcolsep}{3pt} 
\centering
\begin{tabular}{||c||c|c||}
    \hline
    \textbf{Parameter} & \textbf{\textsc{Pythia} Value} & \textbf{\textsc{Herwig} Value} \\
    \hline
    PDF & LHAPDF6:CT09MCS & CTEQ6L1 \\
    \hline
    softQCD & on & on \\
    \hline
    $\alpha_s$ & 0.130 & 0.1185 \\
    \hline
    DLmode & - & 3 \\
    \hline
    pT0Ref / pTmin0 & 2.742289 & 2.87 \\
    \hline
    $\epsilon$ & 0.238 & 0.31 \\
    \hline
\end{tabular}
\caption{Comparison of \textsc{Pythia} and \textsc{Herwig} parameter settings tuned with LHCb experiment.}
\label{lhcb_tunedsettings}
\end{table}

\begin{figure*}[h]
    \centering
    \begin{subfigure}[b]{0.45\linewidth}
        \centering
        \includegraphics[width=\linewidth]{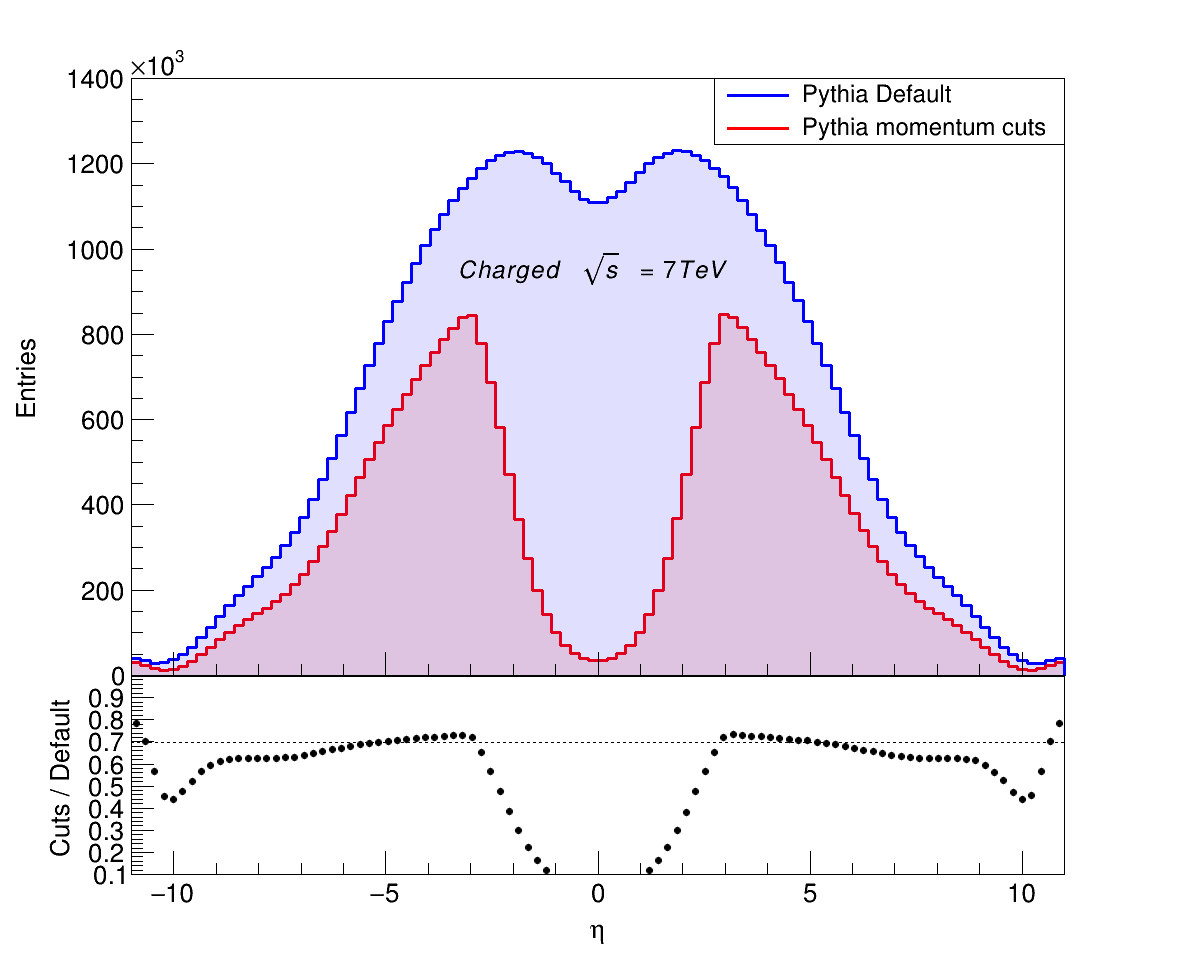}
        \caption{}
    \end{subfigure}
    \hspace{0.01\linewidth} 
    \begin{subfigure}[b]{0.45\linewidth}  
        \centering 
        \includegraphics[width=\linewidth]{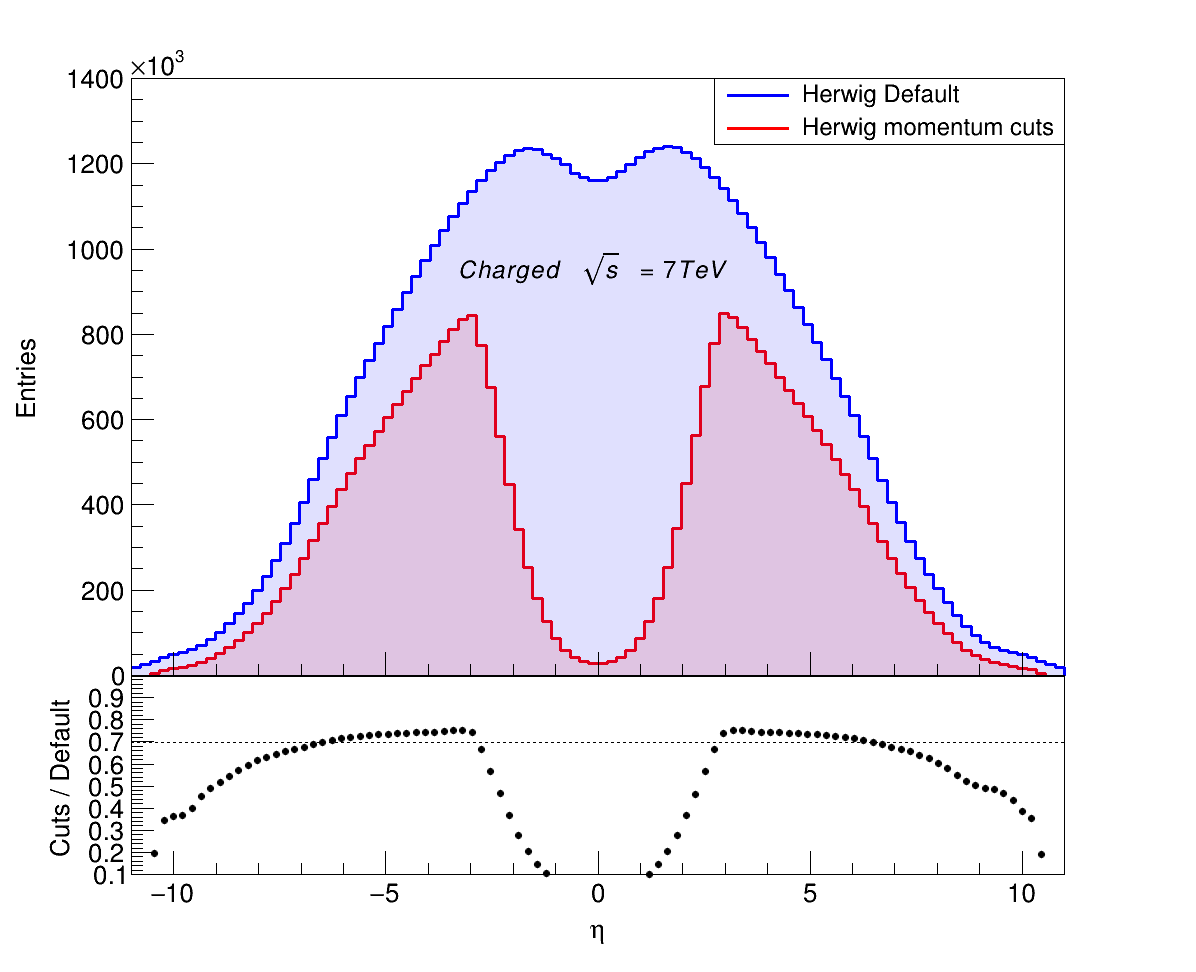}
        \caption{}
    \end{subfigure}
    \caption{Pseudorapidity distributions for charged hadrons at $\sqrt{s}$ = 7 TeV for \textsc{Pythia} in (a) and \textsc{Herwig} in (b), shown with and without momentum requirements. The comparison highlights the effect of momentum cut $>$ 2 GeV/c and transverse momentum cut $>$ 0.2 GeV/c in the generation of charged hadrons in generators.}
    \label{momentum_cuts_dist}
\end{figure*}

For the $\eta$ distribution in Fig.~\ref{all_figs} (a), \textsc{Herwig} underestimates the data for the entire acceptance $\eta$ range, whereas \textsc{Pythia} underestimates the data at low $\eta$ ranges, and overestimates at high $\eta$ ranges. When this distribution is compared to Fig.~\ref{pyth_herwig_comparison_eta7}, there is a similar trend observed between the two MC models.  Fig.~\ref{pyth_herwig_comparison_eta7} seems to be slightly below unity, since there is no cut applied on $p_{T}$ and the momentum of the analysis that could be responsible for the difference. The pseudorapidity distribution with and without the momentum cuts are shown in Fig.~\ref{momentum_cuts_dist} for \textsc{Pythia} in (a) and \textsc{Herwig} in (b). For the $p_{T}$ distribution,  \textsc{Pythia} is in good agreement with the data, whereas \textsc{Herwig} again underestimates the data. This shows a fairly similar pattern to Fig.~\ref{pyth_herwig_comparison_pt7} (a), which is the charged distribution of hadrons with respect to the transverse momentum. If we could expand these plots within the same $p_{T}$ regions and apply the same cuts, one could observe a similar pattern of \textsc{Pythia} dominating the \textsc{Herwig} distribution. Lastly, for the full range charged particle distribution in Fig.~\ref{all_figs} (c), it is quite evident from the plot that \textsc{Pythia} matches the data better than \textsc{Herwig}, making \textsc{Pythia} a better choice of MC generator to be used at LHCb. The results validate the consistency of the MPI model at LHC energies in the forward region, with no unexpected behavior observed.

\section{Verification of Pythia8 parameters for the LHCb Simulation }

An essential ingredient in achieving accurate event simulation is the tuning of Monte Carlo generators to experimental data. 
A verification campaign of the Pythia LHCb-default tuning, obtained from tuning of earlier versions, was carried out for version 8.244. It focused on parameters sensitive to soft-QCD dynamics, such as hadronization, color reconnection, and multiparton interactions. Using measurements of charged-particle production and strange hadron yields in the forward region at $\sqrt{s}=7$ TeV, the parameters were systematically varied and optimized against observables in LHCb. The study was carried out with the  \textsc{Professor} tuning tool \cite{professor} and \textsc{Rivet} \cite{rivet} plugins of relevant LHCb measurements.

\begin{figure*}[h]
    \centering
    \begin{subfigure}[b]{0.45\linewidth}
        \centering
        \includegraphics[width=\linewidth]{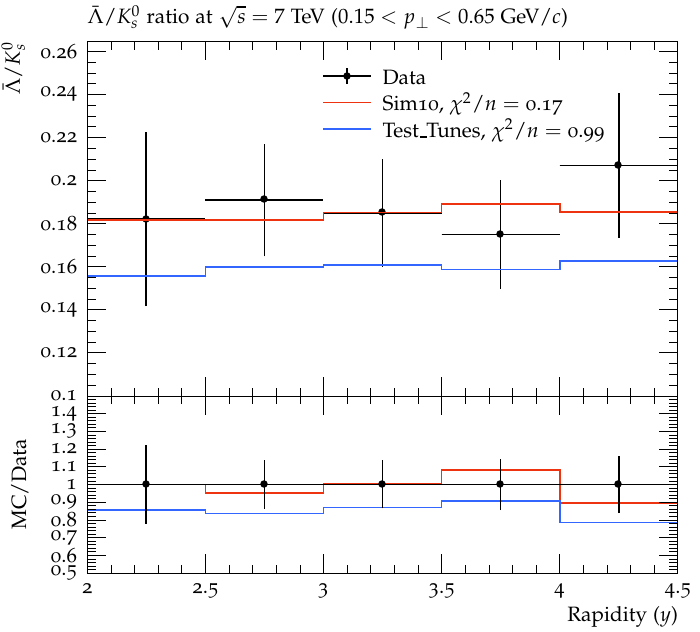}
        \caption{}
    \end{subfigure}
    \hspace{0.01\linewidth} 
    \begin{subfigure}[b]{0.45\linewidth}  
        \centering 
        \includegraphics[width=\linewidth]{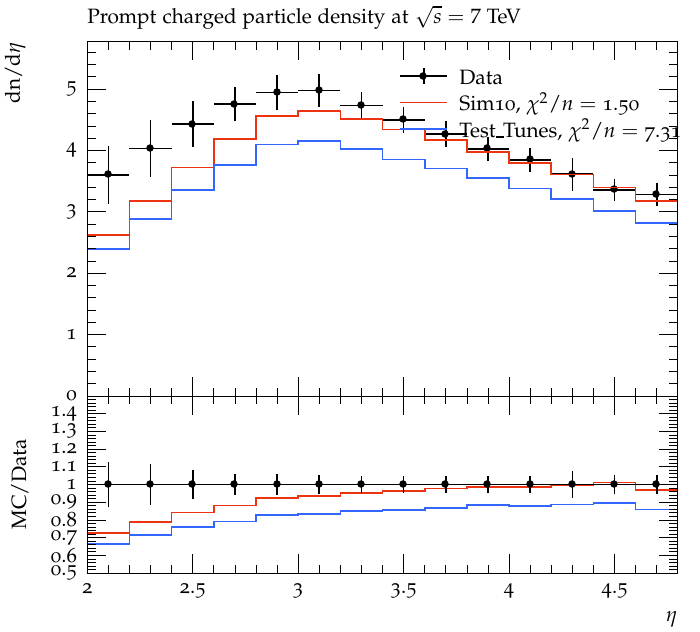}
        \caption{}
    \end{subfigure}
    \caption{Comparison of the LHCb-default tune (Sim10) and Test-Tunes (tuned) at $\sqrt{s}=7$ TeV: (a) $\Lambda/K_{S}$ rapidity ratio, (b) prompt charged-particle density versus pseudorapidity for \textsc{Pythia}. The corresponding $\chi^{2}$ values are indicated.}
    \label{tuning}
\end{figure*}

Figure~\ref{tuning} shows two representative comparisons between the current LHCb- default tune, and the Test-Tunes obtained during this campaign. The $\Lambda/K_{S}^{0}$ ratio as a function of rapidity \cite{rivet_V0} (a) and the pseudorapidity density of prompt charged particles \cite{rivet_prompt} (b) indicates that the LHCb-default tune provides a better agreement with the data, with significantly lower $\chi^{2}/n$ values than the Test-Tunes.

These results emphasize that while exploratory test tunes are valuable for probing sensitivity to different generator parameters, the official LHCb tune remains the most reliable configuration for describing forward-region observables at the LHC. 

\section{Summary and conclusions}

This paper examines the role of Monte Carlo event generators in simulating high-energy particle collisions like those observed at the Large Hadron Collider (LHC). These sophisticated computational tools are essential in both theoretical and experimental studies, allowing researchers to investigate complex phenomena such as particle production, the internal structure of protons, and the fundamental dynamics of particle interactions. Among the most widely used event generators in collider physics are \textsc{Pythia}, \textsc{Herwig}, and \textsc{Sherpa}, which are known for their versatility and general-purpose applications in simulating particle collisions.

This paper focuses specifically on \textsc{Pythia} and \textsc{Herwig}, two of the most prominent event generators used to model LHC data. Both generators are crucial for providing insight into the behavior of particles produced during high-energy collisions, but their performance can vary significantly depending on the choice of parameters. This paper examines how specific parameters affect the particle production process, emphasizing the importance of selecting parameters carefully to obtain accurate results. Two MPI parameters are examined in detail to understand their impact on particle production distributions. It is observed that even slight adjustments in the values of these parameters can lead to substantial changes in multiplicity distributions and energy plots, emphasizing the sensitivity of these models to their input parameters.

A key aspect discussed in this paper is the need for calibration to experimental data, as both \textsc{Pythia} and \textsc{Herwig} are highly sensitive to the chosen parameters. This sensitivity is particularly important in the context of the LHCb experiment, a specialized detector with a forward rapidity coverage at the LHC that focuses on studying rare decays and the physics of heavy quarks (such as b- and c-quarks). Given the specific needs of the LHCb experiment, the event generators used for simulations must be well-tuned to match experimental observations.

This paper also compares the performance of \textsc{Pythia} and \textsc{Herwig} in terms of their accuracy. It is found that \textsc{Pythia} tends to provide a closer match to LHCb data for MPI simulations, making it the preferred choice for simulations in this particular experimental setup. In contrast, while \textsc{Herwig} can also produce valid results, within the limits of the existing available tunes and without a fully uniform tuning effort, \textsc{Pythia} achieves closer agreement with LHCb data under default and LHCb-tuned settings. However, this should not be interpreted as a fundamental shortcoming of \textsc{Herwig} but rather a reflection of the relative level of optimization for the specific dataset under study.

Overall, this paper discusses the critical importance of choosing appropriate parameters and calibrating Monte Carlo event generators to experimental data. It demonstrates that, for the LHCb environment at 7 TeV, \textsc{Pythia} provides better data-to-MC agreement under existing tunes, while emphasizing that systematic tuning efforts are essential to achieve comparable precision across different generators.

\section*{Acknowledgements}
This work is partially supported by the NCN grant UMO-2019/35/O/ST2/00546.

We thank the LHCb Simulation Project for providing and maintaining the software tools and frameworks that supported this work.

\bibliographystyle{spphys}       
\bibliography{template-epjplus}

\end{document}